\documentclass{article}
\usepackage{arxiv}

\usepackage{times}
\usepackage{epsfig}
\usepackage{graphicx}
\usepackage{amsmath}
\usepackage{amssymb}

\usepackage{array}
\usepackage{multirow}
\usepackage{amsmath}
\usepackage{subfigure}
\usepackage{nicefrac}
\usepackage{caption}
\usepackage{color,soul}
\usepackage{eqnarray,amsmath}
\usepackage{algorithm}
\usepackage{algpseudocode}

\usepackage{framed} 
\usepackage{color,soul}
\usepackage{rotating}
\usepackage[normalem]{ulem}

\newcommand{\alert}[1]{\textcolor{red}{#1}}

\usepackage{arydshln}
\usepackage{url}
\title{Beat Detection and Automatic Annotation of the Music of Bharatanatyam Dance using Speech Recognition Techniques}

\author{
  Tanwi~Mallick, Partha Pratim Das, Arun Kumar Majumdar \\
  Department of Computer Science and Engineering\\
  Indian Institute of Technology, Kharagpur,
  India, 721302 \\
  \texttt{tanwimallick@gmail.com, ppd@cse.iitkgp.ac.in, akmj@cse.iitkgp.ac.in} \\
}

\begin{document}
\maketitle

\begin{abstract}
{\em Bharatanatyam}, an Indian Classical Dance form, represents the rich cultural heritage of India. Analysis and recognition of such dance forms are critical for preservation of cultural heritage. Like in most dance forms, a {\em Bharatanatyam} dancer performs in synchronization with structured rhythmic music, called {\em Sollukattu}, that comprises instrumental beats and vocalized utterances ({\em bol}s) to create a rhythmic music structure. Computer analysis of {\em Bharatanatyam}, therefore, requires structural analysis of {\em Sollukattu}s. In this paper, we use speech processing techniques to recognize {\em bol}s. Exploiting the predefined structures of {\em Sollukattu}s and the detected {\em bol}s, we  recognize the {\em Sollukattu}. We estimate the tempo period by two methods. Finally we generate a complete annotation of the audio signal by beat marking. For this we also use information of beats detected from the onset envelope of a {\em Sollukattu} signal~\cite{mallick2018characterization}. For training and test, we create a data set for {\em Sollukattu}s and annotate them. We achieve 85\% accuracy in {\em bol} recognition, 95\% in {\em Sollukattu} recognition, 96\% in tempo period estimation, and over 90\% in beat marking. This is the maiden attempt to fully structurally analyze the music of an Indian Classical Dance form and the use of speech processing techniques for beat marking.
\end{abstract}


\section{Introduction}
\label{sec:introduction}

{\em Bharatanatyam}, an {\em Indian Classical Dance} (ICD) form, represents the rich cultural heritage of India. Modeling, analysis, recognition and interpretation of such dance forms are important to preserve intangible cultural heritage by dance transcription and automatic annotation of dance videos, to create dance tutoring systems, to create animation with avatars, and so on. {\em Bharatanatyam} has a complex language in which the dancers communicate to their audience by telling a story through craftily synchronized visual (postures, gestures, and movements), auditory (beats and {\em bol}s or utterances), and textual (narration and lyrics) information.

Like in most dance forms, a {\em Bharatanatyam} dancer performs in sync with structured rhythmic music, called {\em Sollukattu}. Specific rhythms are created in a {\em Sollukattu} with instrumental beats using {\em Tatta Kazhi}\footnote{Traditionally, a beater beats a {\em Tatta Kazhi} (wooden stick) on a {\em Tatta Palahai} (wooden block) for the instrumental sound and speaks out {\em bol}s like {\em tat}, {\em tei}, {\em ta} etc. as distinct vocalizations of the rhythm.}, {\em Mridangam} etc. Monosyllabic vocal utterances, called {\em bol}s, often accentuate the beats and serve as cues for actions of the dancer. Vocal music may also be used for embellishment. Analysis of a {\em Bharatanatyam} performance, therefore, is critically dependent on understanding the structures of {\em Sollukattu}s.

In this paper, we attempt to automatically annotate the audio signal of a {\em Sollukattu} by detecting its beats, recognizing the accompanying {\em bol}s, recognizing the {\em Sollukattu}, estimating its tempo period, and marking the time-stamps of the beats on the signal. The annotation may be used to denote interesting events like key postures and elemental motion elements, on the {\em Bharatanatyam} dance video and segmented for further visual analysis. 

To keep the complexity of the problem manageable, we work only with {\em Adavu}s of {\em Bharatanatyam}. An {\em Adavu} is a basic unit of {\em Bharatanatyam} performance comprising well-defined sets of postures, gestures, movements and their transitions, and is typically used to train the dancers. There are 58 {\em Adavu}s commonly used\footnote{There are variations between schools -- we follow {\em Kalakshetra}.} in {\em Bharatanatyam} and each {\em Adavu} is synchronized with one of the 23 {\em Sollukattu}s. 
Analysis of a {\em Sollukattu} is a challenging task because it may contain various sources of noise and there are several events (like full-beat, half-beat, and {\em bol}) in the signal to detect. The generation of the music itself may be imperfect due to the lack of skill of the beater or simple human error and fatigue. So we engage a combination of signal processing and speech recognition techniques for the tasks. 

We start with a brief look into various approaches to similar problems in Sec.~\ref{sec:related_work}. We classify the events for a {\em Sollukattu} in Sec.~\ref{sec:model_sollukattu} and present an ontological model. The problem is formally stated in Sec.~\ref{sec:problem_solution} and the approach to solution is outlined. Sec.~\ref{sec:data_set_annotation} presents the data set and annotation. A recognizer for {\em bol}s is discussed in Sec.~\ref{sec:bol_recognizer} along with the signature of audio signals. Sec.~\ref{sec:sollukattu_recognizer} introduces the signature of a {\em Sollukattu} and then presents a recognizer for {\em Sollukattu}s. Tempo period is estimated in Sec.~\ref{sec:tempo_period_estimation}. Based on the analysis and outputs from the earlier sections, we present an algorithm to mark the beats (annotate) on a {\em Sollukattu} signal in Sec.~\ref{sec:beat_marking}. Finally, we conclude in Sec.~\ref{sec:conclusion}.

\section{Related Work \& Approach of Analysis\label{sec:related_work}}
Music is usually created by various instruments like idiophones (percussion instruments),  membranophones (vibrating membranes), aerophones (wind instruments), or chordophones (stringed instruments). Often it is accompanied by human voice which may either function as an instrument, or render speech in melody and harmony to the underlying music, or both. Music usually has rhythm that defines its pattern in time and comprises pitch of high and low tones. Rhythm defines the way the musical sounds and silences are put together in a sequence and often has regular beats. When accompanied by speech-like vocal line (or song), music usually carries a lyrics composed of verses. While there exist several variations to these notions of music and multitude of more parameters (like dynamics, timbre) to define it; the above is a typical characterization of music used in research on acoustic musical signals. Consequently, research has been focused mainly in two areas: (1) Structural Analysis of Music (beat detection, tempo estimation, beat tracking etc.) and (2) Semantic Analysis of Lyrics (song retrieval, segmentation, labeling, and recognition, genre classification, transcription of lyrics, etc.). We take a brief look into these before putting our work into context.

For structural analysis of music, various algorithms for beat detection, tempo estimation, and beat tracking have been reported. Many of these, like~\cite{davies07},~\cite{dixon07}, and~\cite{beattracking}, work on a common framework where first the onset locations are extracted from a time-frequency or sub-band analysis of the signal by using a filter bank or {\em Fast Fourier Transform} (FFT), and then a periodicity estimation algorithm is employed to determine the rate at which these events occur. There are variants of this approach. For example, Peeters et al.~\cite{peeters2011simultaneous} propose a probabilistic framework for estimation of beat and downbeat locations in an audio by considering the tempo period and meter as input. The variations notwithstanding, {\bf we observe that researchers mostly do not consider the vocal sound, if present in the music, for structural analysis.}

Semantic analysis, on the other hand, is primarily undertaken for songs that may consist of musically relevant sounds by the human voice along with the instrumental sound. For example, Mesaros et al.~\cite{mesaros2010automatic} recognize phonemes and words in the audio to align textual lyrics and to retrieve songs, Cheng et al.~\cite{cheng2009multimodal} process lyrics for extracting semantically meaningful segments, Berenzweig et al.~\cite{berenzweig2001locating} locate singing voice segments in music using a speech recognition system, Goto et al.~\cite{goto2004speech} design for speech completion and spotter interface in a background-music playback system, and Scheirer et al.~\cite{scheirer1997construction} present large selection of signal-level features to discriminate regular speech from music. {\bf Most of these use different speech processing techniques for analysis.}

We intend to perform detailed structural analysis of the {\em Sollukattu}\footnote{The word `{\em sollukattu}' originates from the words {\em sollum} (syllables) and {\em kattu} (speaking). It literally means a rhythmic syllable. Here, we refer to the combined audio of instrumental beats and the vocalization as a {\em Sollukattu}.} signals to cater to the requirements of the dance.
%
Interestingly, a {\em Sollukattu} uses human vocalizations in terms of {\em bol}s\footnote{{\em Bol}, meaning {\em bolna} ({\em to speak}), is a mnemonic syllable to define  {\em taalam}.}, which are speech-like signals, to accentuate the rhythm. Hence it calls for speech processing techniques for structural analysis. {\bf Unlike most other work that use beat analysis for estimating music structure and speech recognition for music classification, we also use speech recognition for structural analysis (beat marking and {\em Sollukattu} recognition)}.


Since {\em Sollukattu}s belong to {\em Carnatic Music}, we briefly refer to the related work in {\em Indian Hindustani \& Carnatic Music}\footnote{Hindustani and Carnatic Music are two main sub-genres of Indian Classical Music. {\em Bharatanatyam} uses Carnatic Music.}
Structural analyses have been used by~\cite{klapuri06},~\cite{klapuri03},~\cite{srinivasamurthy12}, and~\cite{gulati12}, to address the problem of estimating the meter of a musical piece. The two stage comb filter-based approach (originally proposed for double / triple meter estimation) is extended to septuple meter (such as 7/8 time-signature) in~\cite{gulati12}. Its performance is evaluated on a sizable Indian music database. In~\cite{sridhar2009raga}, Sridhar and Geetha propose an algorithm to segment the instrumental and the vocal signals. The frequency components of the signal are determined on the voice signal and  mapped onto the {\em swara}\footnote{{\em Swara}, in Sanskrit, means a note in the successive steps of the octave.} sequence. Srinivasamurthy et al. present an algorithm~\cite{srinivasamurthy12} using a beat similarity matrix and inter-onset interval histogram to automatically extract the sub-beat structure and the long-term periodicity of a musical piece. They achieve 79.3\% accuracy on an annotated {\em Carnatic} music data set. There has, however, been no attempt to structurally analyze the music of ICD.

\section{{\em Sollukattu} -- The Audio of {\em Adavu}s\label{sec:model_sollukattu}}
{\em Sollukattu}s follow rhythmic musical patterns, called {\em Taalam}\footnote{{\em Taalam} is the Indian system for organizing and playing metrical music.}, created by combination of instrumental and vocal sounds to accompany {\em Bharatanatyam Adavu} performances.
A repeated cycle of {\em Taalam} consists of $\lambda$ number of equally spaced beats grouped into combinations of patterns. Time interval between any two beats is always equal and is called the {\em Tempo Period}. 
The specific way the beats are marked 
is determined by the {\em Taalam}. While different {\em Taalam}s are used in {\em Bharatanatyam}, {\em Adi} ($\lambda$ = 8 beats' pattern) and {\em Roopakam Taalam} ($\lambda$ = 6) are most common. Finally, a {\em Taalam} is devoid of a physical unit of time and is acceptable as long as it is rhythmic in some unit. With a base time unit, however, {\em Bharatanatyam} deals with three speeds\footnote{{\em Kaalam}s or {\em Tempo}s are -- Base speed or {\em Vilambitha Laya}, Double (of base) speed or {\em Madhya Laya}, and Quadruple (of base) speed or {\em Duritha Laya}.}  called {\em Kaalam} or {\em Tempo}. 

In a {\em Sollukattu}, instrument and voice both follow in sync to create a pattern of beats consisting of: (1) {\em Instrumental Sub-stream} from instrumental strikes, and (2) {\em Vocal Sub-stream} from vocalizations or {\em bol}s. In {\em Instrumental} and {\em Vocal Sub-streams} of a {\em Sollukattu}, beating and {\em bol}s are created in sync by the {\em beater}. 
To analyze this musical structure we first identify events in it and then formulate a model for it. 

\subsection{Audio Events of {\em Sollukattu}s}
An {\em Event} denotes the occurrence of a {\em Causal Activity} in the audio stream as listed in Tab.~\ref{tbl:events}. An event has:
\begin{table}[!ht]
\caption{List of Events of {\em Sollukattu}s\label{tbl:events}}
\centering
\begin{small}
\begin{tabular}{|l|p{4cm}|p{2cm}|} 
\hline
\multicolumn{1}{|c|}{\bf Event} & \multicolumn{1}{c|}{\bf Description} & \multicolumn{1}{c|}{\bf Label} \\ \hline \hline
 $\alpha^{fb}$ & Full-beat$^1$ or $1$-beat or B with {\em bol} & {\em bol}$^2$, downbeat$^3$, upbeat$^4$ \\
 $\alpha^{hb}$ & Half-beat$^5$ or $\frac{1}{2}$-beat or HB with {\em bol} & {\em bol} \\
 $\alpha^{qb}$ & Quarter-beat$^6$ or $\frac{1}{4}$-beat or QB with {\em bol} & {\em bol} \\
 $\alpha^{fn}$ & $1$-beat having no {\em bol}& upbeat, \newline stick-beat$^7$ ($\bot$)  \\
 $\beta$ & {\em bol} is vocalized & {\em bol} \\ \hline
 \multicolumn{3}{c}{ } \\
\end{tabular}

\begin{tabular}{rp{7cm}}
1: & A beat, often referred to as full-beat or $1$-beat, is the basic unit of time -- an instance on the timescale \\
2: & {\em bol}s accompany some beats ($1$-, $\frac{1}{2}$- or $\frac{1}{4}$-)\\
3: & The first $1$-beat of a bar \\
4: & The last $1$-beat in the previous bar which immediately precedes, and hence anticipates, the downbeat \\
5: & A $\frac{1}{2}$-beat is a soft strike at the middle of a $1$-beat to $1$-beat gap or tempo period \\
6: & A $\frac{1}{4}$-beat is a soft strike at the middle of a $1$-beat to $\frac{1}{2}$-beat or a $\frac{1}{2}$-beat to $1$-beat gap  \\
7: & A stick-beat ($\bot$) has only beating and no {\em bol} \\ \hline
\end{tabular}
\end{small}
\end{table}

\begin{enumerate}
\item {\em Type}: Type relates to the causal activity of an event. 

\item {\em Time-stamp / range}: The time of occurrence of the causal activity of the event. This is elapsed time from the beginning of the stream and is marked by a function $\tau(.)$. Often a causal activity may spread over an interval $[\tau_s, \tau_e]$ which will be associated with the event. 

\item {\em Label}: One or more optional labels may be attached to an event annotating details for the causal activity.
 
\item {\em ID}: Every instance of an event in a stream is distinguishable. These are sequentially numbered in the temporal order of their occurrence. 

\end{enumerate}

\subsection{Ontological Model of {\em Sollukattu}s}
\begin{figure}
\centering
\begin{tabular}{c}
\includegraphics[width=9cm]{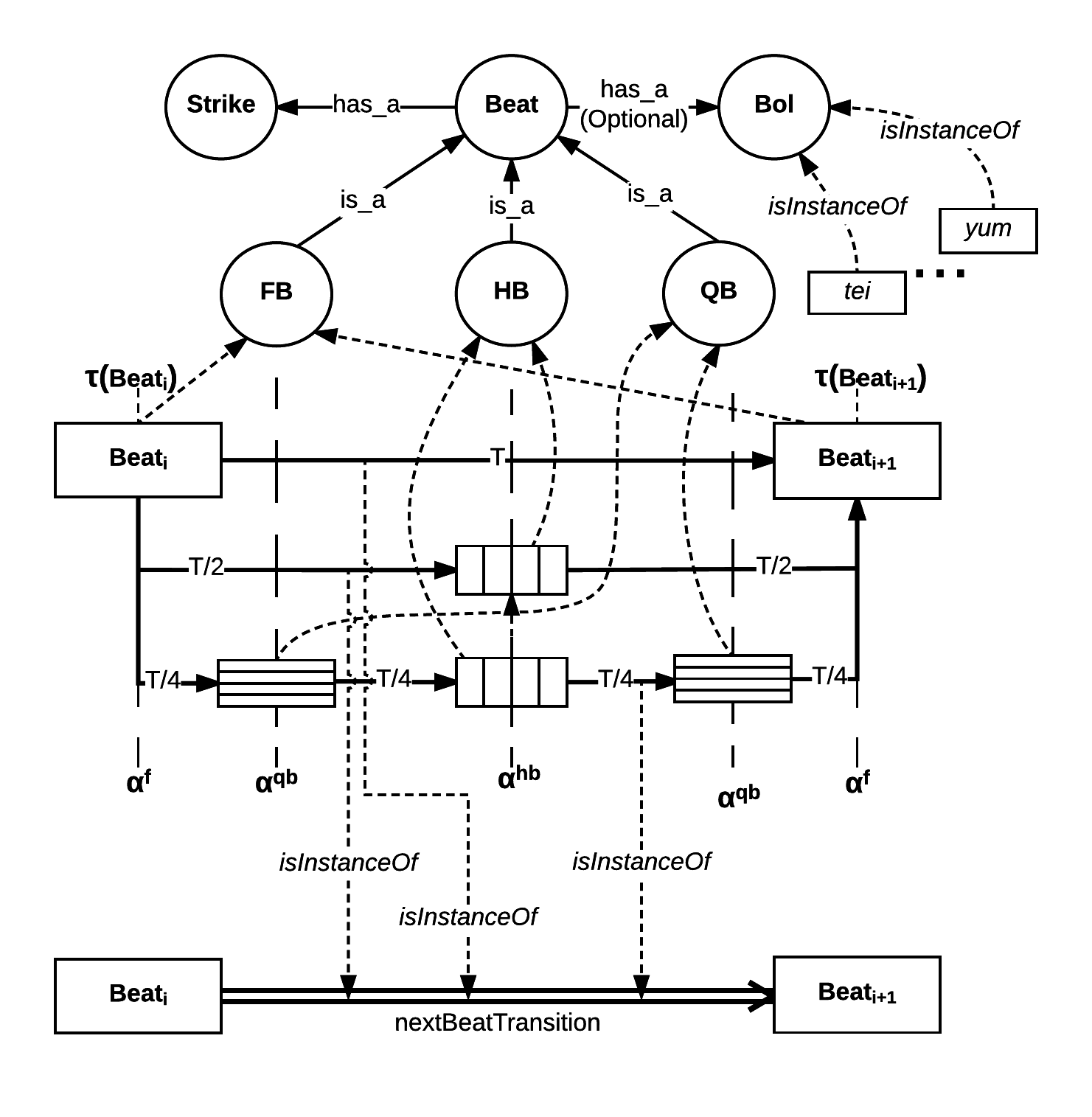} \\
(a) Model of Beats, {\em Bol}s and Transitions \\ \\
\includegraphics[width=9cm]{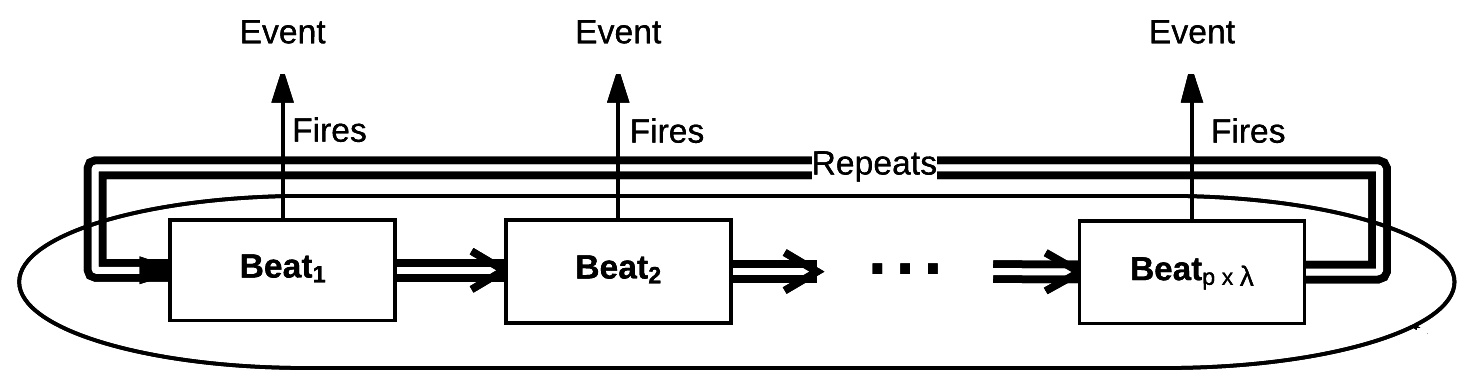} \\
(b) Model of {\em Sollukattu}
\end{tabular}
\caption{Event-Driven Ontological Model of Audio}
\label{fig_ontology_of_audio}
\end{figure}

We present the ontology of a {\em Sollukattu} in Fig.~\ref{fig_ontology_of_audio} highlighting the taxonomy, the partonomy, and the major relationships in the musical structure. The concept classes are shown in ellipses and the instances are marked with rectangles (related by $isInstanceOf$). 

Fig.~\ref{fig_ontology_of_audio}(a) represents the relationships between various types of beats, strikes, and {\em bol}s as discussed above. Using $T$ as tempo period ($1$-beat to $1$-beat gap), we then show the possible transitions ($nextBeatTransition$ relation) between two consecutive $1$-beat instances -- $Beat_i$ and $Beat_{i+1}$. The transition can be of any one of three kinds that either has no intervening $\frac{1}{2}$-beat, or has one $\frac{1}{2}$-beat (vertical bars), or has one $\frac{1}{2}$-beat and one or two $\frac{1}{4}$-beats (horizontal bars). It also marks the events and the time-stamps.

Fig.~\ref{fig_ontology_of_audio}(b) shows that a {\em Sollukattu} is formed of a ($has\_a$) sequence of $p \times \lambda$ number of beat-to-beat transitions by some specialization of $nextBeatTransition$ where $p = 1, 2, 4, 6, 8, \cdots$. This defines the basic rhythmic structure in terms of its {\em Taalam}.  We show the {\em Adi} ($\lambda = 8$) and {\em Roopakam} ($\lambda = 6$) {\em Taalam}s as specializations. A {\em Sollukattu} based on {\em Adi} ({\em Roopakam}) {\em Taalam}, is called a 8-(6-) {\em Recurrent Sollukattu}. 
There are 23 {\em Sollukattu}s in total. 6 of these -- {\em Kartati-Utsanga-Mandi-Sarikkal (KUMS)}, {\em Tatta B \& G}, {\em Tirmana A, B \& C} -- are 6-Recurrent while the rest -- {\em Joining A, B \& C}, {\em Kuditta Mettu}, {\em Kuditta Nattal A \& B}, {\em Kuditta Tattal}, {\em Natta}, {\em Paikkal}, {\em Pakka}, {\em Sarika}, {\em Tatta A, C, D, E \& F}, {\em Tei Tei Dhatta (TTD)} -- are 8-Recurrent. {\em Tirmana A, B \& C} use $p = 2$ while others use $p = 1$.

\section{Problem Statement \& Solution Approach\label{sec:problem_solution}}
Let $S$ be the set of {\em Sollukattu}s. The recording of a {\em Sollukattu} $s \in S$ is a
discrete-time audio signal $f^s(t)$ defined as:
\begin{equation}
f^s \equiv f^s(t) \equiv \{f^s_0, f^s_1, \cdots, f^s_{n-1}\} \label{eqn:audio_signal}
\end{equation}
where $0 \leq t \leq \mathcal{T}$, $\mathcal{T}$ is the duration of the signal, $f^s(t)$ is a sequence $\{f^s_i\}$ of sampled and quantized values, $n = \mathcal{T} \times r$ is the number of samples, and $r$ is the sampling rate.

The sampling and quantization of audio is performed by the recorder at a rate of $r = 44100$ sample /sec. $\approx$ 22.7 $\mu s$ = $\delta_s$. Hence there is a time-stamp every $\delta_s$ sec. available on the audio packets of $f^s$. Thus we can mark $f^s$ in time with $\delta_s$ sec. resolution. However, we deal with coarse-grained events like beats and {\em bol}s that usually span over 100 ms.

Given an audio signal $f^s$, we want to solve for following:
\begin{framed}
\begin{enumerate}
\item {\bf Recognize the {\em Sollukattu} $s$ of $f^s$}
\item {\bf Mark $f^s$ with time-stamps of beats, beat information ($1$, $\frac{1}{2}$ or $\frac{1}{4}$) and associated {\em bol}s}
\end{enumerate}
\end{framed}

This, in turn, needs the solution of the following:
\begin{enumerate}
\item Recognize and build the sequence of {\em bol}s in $f^s$ (Sec.~\ref{sec:bol_recognizer})
\begin{itemize}
\item To detect {\em bol}s ($\beta$ events), we first segment $f^s$ using silence intervals. The  MFCC ({\em Mel Frequency Cepstral Coefficients}) features of segmented non-silent slices are used with {\em Gaussian Mixture Model} (GMM) to classify the {\em bol}s. The signal $f^s$ is then represented in terms of a string signature (called, {\em Signal Signature}) comprising the recognized {\em bol}s. 
\end{itemize}

\item Recognize the {\em Sollukattu} $s$ of $f^s$ (Sec.~\ref{sec:sollukattu_recognizer})
\begin{itemize}
\item To recognize the {\em Sollukattu}s, we build a dictionary of string signatures of {\em bol}s (called, {\em Sollukattu Signature}) for every {\em Sollukattu}. We match the {\em Signal Signature} of $f^s$ with the  {\em Sollukattu Signature}s in the dictionary using an edit distance.
\end{itemize}

\item Estimate the tempo period $T^s$ of $s$ from $f^s$ (Sec.~\ref{sec:tempo_period_estimation})
\begin{itemize}
\item We estimate the tempo period using two methods:
\begin{enumerate}
\item Working directly with the signal $f^s$, we estimate the tempo period by Comb (resonating) filter 
\item We estimate the tempo period from the Longest Common Sub-string (LCS) between {\em Signal Signature} and {\em Sollukattu Signature}
\end{enumerate}
\end{itemize}

%
\item Mark time-stamps and {\em bol}s of beats in $f^s$ (Sec.~\ref{sec:beat_marking})
\begin{itemize}
\item Beat Positions are a sequence of time-stamps $\tau_i, i \geq 0$ on $f^s$. For tempo period $T^s$ of $s$, if there is a $1$-beat ($\alpha^f$ event) at $\tau_i$, we have $\tau_{i+1} - \tau_i \approx T^s, \forall i$. For a $\frac{1}{2}$-beat ($\alpha^h$ event), $\tau_{i+1} - \tau_i \approx T^s/2$. In~\cite{mallick2018characterization} we detect beats using detection of onsets in $f^s$ and subsequent refinement of the set of detected onsets. This, however, works only for $1$-beats. 
\item We use the information of detected beats, detected {\em bol}s and estimated tempo period to design an algorithm that traverses on $f^s$ and marks various events, time-stamps and {\em bol}s on the signal by exploiting the structural properties of a {\em Sollukattu}. 
\end{itemize}

%
\end{enumerate}

\section{Data Set \& Annotation\label{sec:data_set_annotation}}
No data set for {\em Sollukattu}s are available for training and testing purposes of the research. Hence, we had to create a data set by recording performances and then annotating them
with the help of expert {\em Bharatanatyam} dancers. A part of the data set (SR1) has been published in {\em Audio Data (Sollukattu)}~\cite{mallick2017annotatedAudioData} for reference and use by researchers.

\subsection{Data Set}
We record 6 sets (Tab.~\ref{tbl:Sollukattu_Data_Set}) of 23 {\em Sollukattu}s using a {\em Zoom H2N Portable Handy Recorder}. The first of the sets (SR1) was recorded for only a single bar (cycle) while others are done for 4 bars. Also, in SR1 and SR2, few {\em Sollukattu}s are recorded multiple times. 162 {\em Sollukattu} files corresponding to recording sets SR1--SR6 (Tab.~\ref{tbl:Sollukattu_Data_Set}) have been recorded and subsequently annotated as follows.


\begin{table}[!ht]
\caption{{\em Sollukattu} Data Set as recorded\label{tbl:Sollukattu_Data_Set}}
\centering
\begin{small}
\begin{tabular}{|l|l|r|r|r|} \hline 
\multicolumn{1}{|c}{\bf Recording} & 
\multicolumn{1}{|c}{\bf Beater}  & 
\multicolumn{1}{|c}{\bf \# of}		& 
\multicolumn{1}{|c}{\bf \# of} &
\multicolumn{1}{|c|}{\bf \# of} \\ 

\multicolumn{1}{|c}{\bf Set \#} & 
\multicolumn{1}{|c}{\bf  \#}  & 
\multicolumn{1}{|c}{\bf {\em Sollukattu}s} & 
\multicolumn{1}{|c}{\bf Cycles}	&
\multicolumn{1}{|c|}{\bf Recordings} \\ \hline \hline
SR1	&	Beater 1	&	23	&	1	&	30	\\ \hline
SR2	&	Beater 1	&	23	&	4	&	40	\\ \hline
SR3	&	Beater 1	&	23	&	4	&	23	\\ \hline
SR4	&	Beater 1	&	23	&	4	&	23	\\ \hline
SR5	&	Beater 2	&	23	&	4	&	23	\\ \hline
SR6	&	Beater 3	&	23	&	4	&	23	\\ \hline
\multicolumn{4}{l}{\bf Total} & \multicolumn{1}{r}{\bf 162} \\
\multicolumn{5}{p{8cm}}{\footnotesize $\bullet$ In SR1 and SR2, a few {\em Sollukattu}s are recorded multiple times}
\end{tabular}
\end{small}
\end{table}

\subsection{Annotation}\label{sec:annotations}
Annotation of a {\em Sollukattu} involves the following:

\begin{enumerate}
\item Identification the beats and marking them on the signal. Every beat should be marked with its time-stamp as a $1$-, $\frac{1}{2}$-, $\frac{1}{4}$-, or stick beat.
\item Identification the {\em bol}s and their associations with beats.
\item Marking parts of the signal that are silent.
\item Estimation of the tempo period.
\item Marking of the bars and determination of the number of beats in a bar.
\item Documentation of the annotations in Excel.
\end{enumerate}

\begin{figure}[!ht]
\centering
\includegraphics[width=8.75cm]{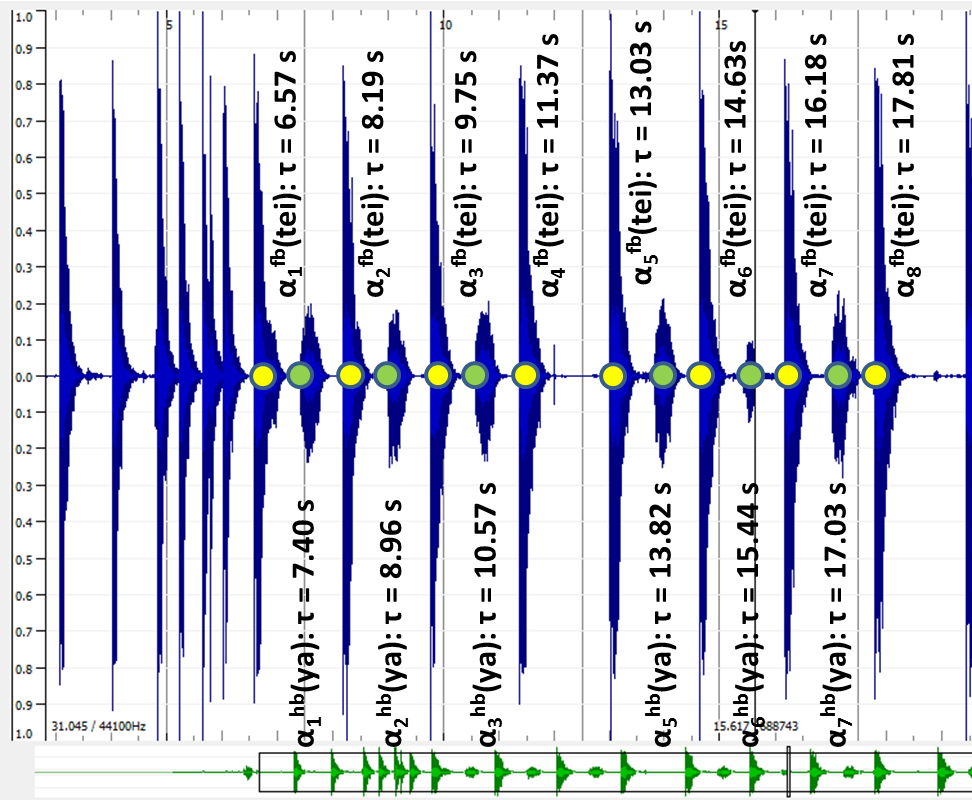} 

\centering
\begin{footnotesize}
\begin{tabular}{p{8.75cm}}
$\bullet$ No. of bars = 2, $\lambda$ = 8 and $T = 1.56$ sec. $1$-beats ($\alpha^{fb}$, yellow) and $\frac{1}{2}$-beats ($\alpha^{hb}$, green) are highlighted with {\em bol}s and time-stamps (Tab.~\ref{tbl:Sollukattu_bols_Tatta_C}). 
\end{tabular}
\end{footnotesize}
\caption{Annotations of beats and {\em bol}s for {\em Tatta C Sollukattu}
\label{fig:Tatta_C_Annotations}}
\end{figure}

Typical annotations are illustrated in Tab.~\ref{tbl:Sollukattu_bols_Tatta_C}  and Fig.~\ref{fig:Tatta_C_Annotations}. We often write the {\em bol}s of a {\em Sollukattu} as a sequence, grouping the {\em bol}s of the same tempo period with [] brackets. Hence, for the {\em Tatta C Sollukattu} (Tab.~\ref{tbl:Sollukattu_bols_Tatta_C}), the {\em bol} sequence is: 

$[tei\ ya] [tei\ ya] [tei\ ya] [tei] [tei\ ya] [tei\ ya] [tei\ ya] [tei]$.
\begin{table}[!ht]
\caption{Annotations of {\em Tatta C Sollukattu} (Fig.~\ref{fig:Tatta_C_Annotations}) \label{tbl:Sollukattu_bols_Tatta_C}}
\centering
{\renewcommand{\arraystretch}{1.3}%
\begin{footnotesize}
\begin{tabular}{|l|r|r|r||l|r|r|r|} \hline
\multicolumn{1}{|c|}{\bf Event}	& \multicolumn{1}{c|}{\bf Time} & \multicolumn{1}{c|}{\bf B} & \multicolumn{1}{c||}{\bf HB} & \multicolumn{1}{c|}{\bf Event}	& \multicolumn{1}{c|}{\bf Time} & \multicolumn{1}{c|}{\bf B} & \multicolumn{1}{c|}{\bf HB} \\ 

\hline \hline
\textcolor{blue}{$\alpha^{fb}_1$(tei)}		&		6.57	&	\multicolumn{1}{r|}{}	&		&	\textcolor{blue}{$\alpha^{fb}_5$(tei)}		&		13.03	&		\multicolumn{1}{r|}{1.66} 	&		\\ \hline
\textcolor{blue}{$\alpha^{hb}_1$(ya)}	&		7.40	&	\multicolumn{1}{r|}{}	&		0.83	 	&	\textcolor{blue}{$\alpha^{hb}_5$(ya)}		&		13.82	&	\multicolumn{1}{r|}{}	&		0.79			\\ \hline
\textcolor{red}{$\alpha^{fb}_2$(tei)}		&	8.19	&		\multicolumn{1}{r|}{1.62} 	&	 	&	\textcolor{red}{$\alpha^{fb}_6$(tei)} 	&		14.63	&		\multicolumn{1}{r|}{1.60} 	&		\\ \hline
\textcolor{red}{$\alpha^{hb}_2$(ya)}		&		8.96	&	\multicolumn{1}{r|}{}	&		0.77		&	\textcolor{red}{$\alpha^{hb}_6$(ya)} 		&		15.44	&	\multicolumn{1}{r|}{}	&		0.81			\\ \hline
\textcolor{blue}{$\alpha^{fb}_3$(tei)}		&		9.75	&		\multicolumn{1}{r|}{1.56} 	&		&	\textcolor{blue}{$\alpha^{fb}_7$(tei)} 	&		16.18	&		\multicolumn{1}{r|}{1.55} 	&		\\ \hline
\textcolor{blue}{$\alpha^{hb}_3$(ya)}		&		10.57	&	\multicolumn{1}{r|}{}	&		0.82	&	\textcolor{blue}{$\alpha^{hb}_7$(ya)}		&		17.03	&	\multicolumn{1}{r|}{}	&		0.85			\\ \hline
\textcolor{red}{$\alpha^{fb}_4$(tei)}		&		11.37	&		\multicolumn{1}{r|}{1.62} 	&		&	\textcolor{red}{$\alpha^{fb}_8$(tei)} 	&		17.81	&		\multicolumn{1}{r|}{1.63} 	&		\\ \hline
\end{tabular}
\end{footnotesize}
}
\end{table}

\section{{\em Bol} Recognition}\label{sec:bol_recognizer}

\begin{figure}
\centering
\includegraphics[width=9cm]{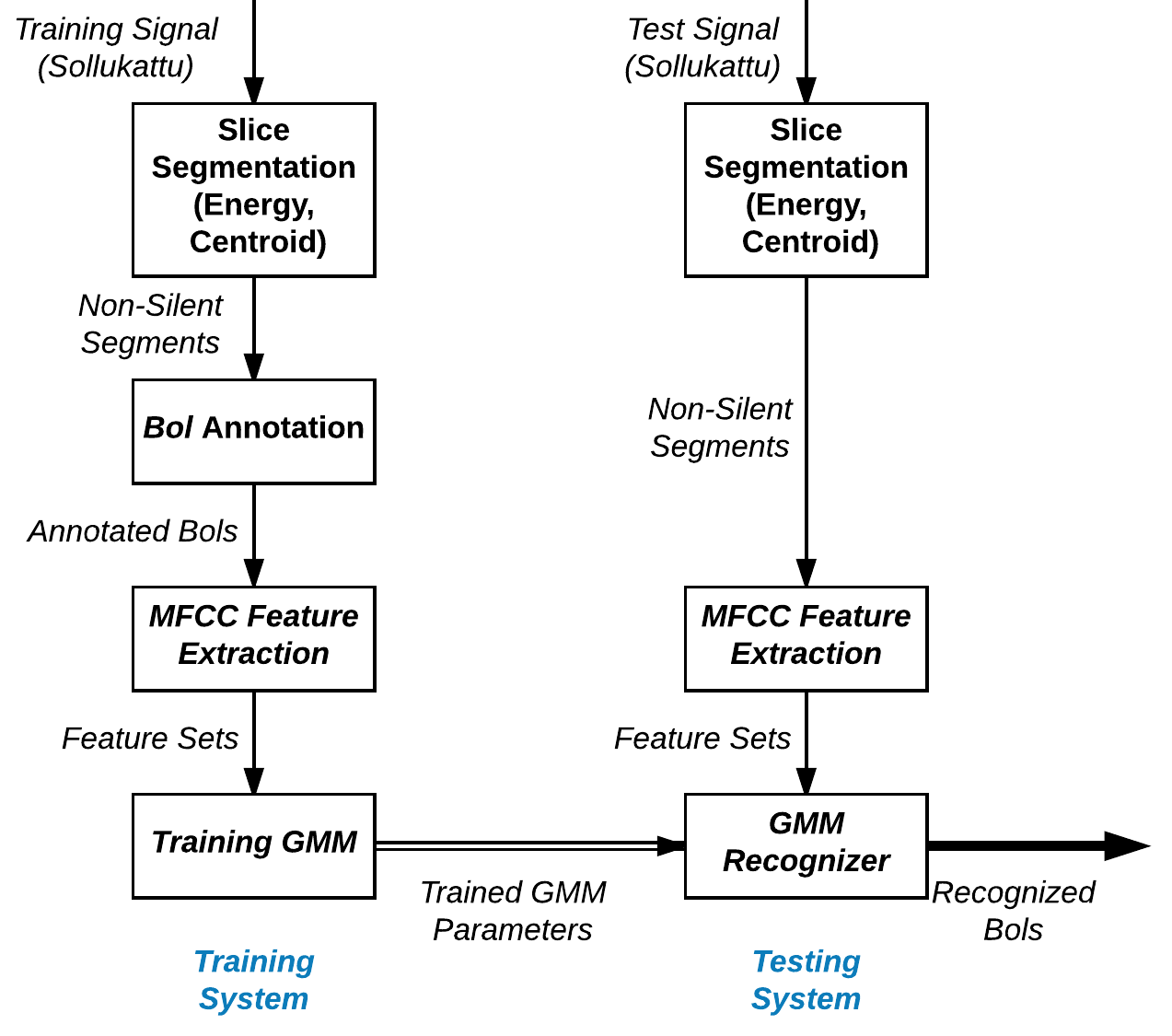}

\caption{Overview of the {\em Bol} Recognition System}
\label{fig_bol_recognition}
\end{figure}

The {\em bol} recognition system is shown in Fig.~\ref{fig_bol_recognition}. In order to train the system, we manually segment the audio signal of every {\em Sollukattu} by removing the silence parts (during annotation). This step generates a number of segmented signals of {\em bol}s and {\em stick-beat}s. We collect the segments from all {\em Sollukattu} signals in the training set and annotate every segment with the {\em bol} class. MFCC features are extracted for the {\em bol} classes and a GMM classifier is trained. 

For testing, we segment the test {\em Sollukattu} audio signal (Sec.~\ref{sec:audio_segmentation}), extract MFCC features for every segment, and then recognize every {\em bol} using a GMM. Finally, we build a sequence of recognized {\em bol}s into a signature (Sec.~\ref{sec:signal_signature}). 

\subsection{Segmentation of Audio Signal into {\em Bol} Segments} \label{sec:audio_segmentation}
To segment $f^s \equiv f^s(t)$ into individual {\em bol} signals, we detect the silence periods in $f^s$ (having value very close to zero) and segment it into a sequence of non-silent slices. A {\em bol} (or {\em stick-beat}) can only be a non-silent slice of signal. A non-silent slice $\hat{f}^s_i(t)$ is non-zero in the interval $\tau(\hat{f}^s_i(t)) = [\tau_{s_i},\tau_{e_i}]$ and is (almost) zero elsewhere. It is defined as:

\begin{center}
\begin{tabular}{llll} 
$\hat{f}^s_i(t)$ & = & $0$, & $0 \leq t < \tau_{s_i}$ \\
 & = & $f^s(t)$, & $\tau_{s_i} \leq t \leq \tau_{e_i}$ \\
 & = & $0$, & $\tau_{e_i} < t \leq \mathcal{T}$ \\ 
\end{tabular}
\end{center}

The signal $f^s$, approximated in terms of non-silent slices, is:
\begin{equation}
f^s(t) \approx \sum_{i=0}^{k^s-1} \hat{f}^s_i(t) \label{eqn:silence_segments}
\end{equation}
where $\tau(\hat{f}^s_i(t)) = [\tau_{s_i},\tau_{e_i}]$, $\forall i, 0 \leq i < k^s-1$, $k^s$ is the number of non-silent slices in $f^s(t)$, and $0 \leq \tau_{s_i} < \tau_{e_i} < \tau_{s_{i+1}} < \tau_{e_{i+1}} \leq \mathcal{T}$. That is, the non-zero parts of the non-silent slices are non-overlapping. Typically, $k^s \gg \lambda^s$, number of $1$-beats in a bar of {\em Sollukattu} $s$. Hence, the signal has lot more beats or events than a bar of a {\em Sollukattu} $s$. Ignoring the silence periods, $f^s(t)$ can be expressed by a sequence of slices as:
\begin{equation}
f^s(t) \approx\  <\hat{f}^s_i(t)\ |\ 0 \leq i < k^s -1> \label{eqn:non_silent_slices}
\end{equation}
Every $\hat{f}^s_i(t)$ represents a {\em bol} or {\em stick-beat}, that is, $\alpha^{fb}$, $\alpha^{hb}$, or $\alpha^{fn}$ events. To convert $f^s(t)$ into the sequence of non-silent slices as above, we use the {\em signal energy} and the {\em spectral centroid} (\cite{giannakopoulos2009study}) of the audio signal for silence removal and segmentation because the energy of the mixed sound (voice and instrumental) is expected to be larger than the energy of the silent segments. 


\subsubsection{Segmentation by Silence Detection}\label{sec:silence_removal}
We divide the signal $f^s(t)$ into overlapping short-term frames, each having a time window $win = 0.090$ sec., to compute the silence / non-silence periods. The overlap is taken as $step = 0.010$ sec. Thus the $i^{th}$ frame has $N = win * \delta_s$ samples given by $x_i(n), n=1,...,N$ where $\delta_s = 44100$ sec$^{-1}$ is the sampling rate. Signal energy ($E_i$) and spectral centroid ($C_i$) features are then calculated for every frame as (1) {\em Signal Energy}, $E_i = \sum_{n=1}^{N}x^2_i(n)/N$, and (2) {\em Spectral Centroid}, $C_i = (\sum_{n = 1}^N (n+1)X_i(n)) / (\sum_{n=1}^N X_i(n))$, where $X_i(n)$, $n = 1,...,N$ are the DFT coefficients of samples of the $i^{th}$ frame.

\begin{figure}[!ht]
 \centering
	\includegraphics[width=9cm]{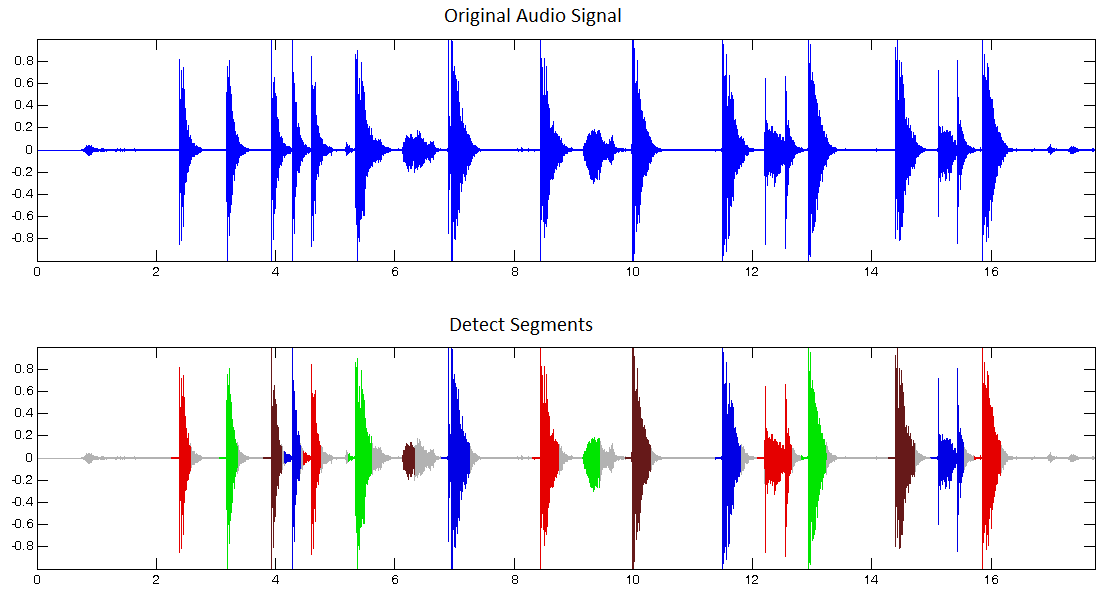}
	
	\begin{footnotesize}
	\begin{tabular}{l}
	$\bullet$ Silence parts are shown in grey \\
	$\bullet$ Different {\em bol} segments are shown in different colors 
	\end{tabular}
	\end{footnotesize}
	\caption{Silence Removal and Segmentation \label{fig:SilenceDetection}}
\end{figure}

The sequence of frames are now converted to sequence of feature values. We threshold based on feature values to remove the silence parts in the audio signal. The thresholds are computed in 3 steps: (1) Compute the histogram of the values in the feature sequence, (2) Detect the local maxima of the histograms, and (3) Let $M_1$ and $M_2$ be the positions of the first and second local maxima respectively. The threshold value is computed as $T = (W * M_1 + M_2)/(W+1)$
where $W$ is a weight to control the cut-off. This process is performed for both feature sequences, leading to two thresholds: $T_E$ (energy) and $T_C$ (spectral centroid). For the $i^{th}$ frame, if either $E_i < T_E$ or $C_i < T_C$ holds, it is considered a part of the silent segment and marked accordingly. The sequence of the remaining frames form the non-silent segments $\hat{f}^s_i(t), 0 \leq i < k^s$, each representing a {\em bol} or a {\em stick-beat}. For every segment $\hat{f}^s_i(t)$ we also mark the start and the end times $\tau(\hat{f}^s_i(t))$ from the frames involved in it. In Fig.~\ref{fig:SilenceDetection}, we illustrate the silence detection for an audio signal. 

We use $W = 0$. As a number of {\em Sollukattu}s have $1$- as well as $\frac{1}{2}$-beats, and as the beating is relatively weak for a $\frac{1}{2}$-beat (though the vocalization is done with the same loudness), the two maxima $M_1$ and $M_2$ in the histograms correspond to the energy from $1$- and $\frac{1}{2}$-beats respectively. Hence the use of $W > 0$ removes part of the $\frac{1}{2}$-beat and weakens the corresponding {\em bol} signal.

\subsection{Signature of Audio Signal}\label{sec:signal_signature}
Let $G$ be a {\em bol recognizer} that takes a non-silent signal slice and recognizes the {\em bol}:
\begin{equation}
G\ :\ \mathcal{F} \rightarrow \mathcal{B} \cup \{\top\} \label{eqn:bol_recognizer}
\end{equation}
where $\mathcal{F}$ is a set of slices $\hat{f}^s_i(t)$, $0 \leq i < k$, $k = |\mathcal{F}|$ is the number of slices in $\mathcal{F}$, $\mathcal{B}$ is the set of {\em bol}s, $\top$ denotes Undefined / Unrecognized {\em bol} arising from {\em stick-beat} (or from some failed recognition). So a {\em bol} event $\hat{\beta_i}$ is recognized as:
$$G(\hat{f}^s_i(t)) = \hat{\beta_i} \in \mathcal{B} \cup \{\top\}$$
with {\em bol} event time interval $\tau(\hat{\beta_i}) = \tau(\hat{f}^s_i(t)) = [\tau_{s_i},\tau_{e_i}]$. We use $\tau(\hat{\beta_i}) = \tau_{s_i}$ or $\tau(\hat{\beta_i}) = (\tau_{s_i}+\tau_{e_i})/2$ when we need one instant in place of an interval. Repeatedly using $G$ (Eqn.~\ref{eqn:bol_recognizer}) on every slice of an audio signal $f^s(t)$ (Eqn.~\ref{eqn:non_silent_slices}), we get a sequence $\Gamma(f^s)$ of recognized {\em bol}s for $f^s(t)$ as:
\begin{equation}
\Gamma(f^s) = <\hat{\beta_i}\ |\ 0 \leq i < k, \hat{\beta_i} \in \mathcal{B}> \label{eqn:signal_signature}
\end{equation}

$\Gamma(f^s)$ is called the {\em Signal Signature} of $f^s(t)$. While building the sequence $\Gamma(f^s)$ we drop the unrecognized symbol, $\top$. We use $\Gamma(f^s)$ later for recognizing {\em Sollukattu}s. 

\subsection{{\em Bol} Recognizer}\label{sec:bol_segmentation}
To build $G$, we first construct a {\em Bol} vocabulary comprising all {\em bol}s in $\mathcal{B}$, use MFCC features to represent every {\em bol} and engage {\em Gaussian Mixture Model} (GMM) as a recognizer.

\subsubsection{{\em Bol} Vocabulary}
The {\em bol} vocabulary, as identified by the experts, is given in Tab.~\ref{tbl:bol_reco_data}. We also denote a class for {\em stick-beats} or $\bot$ (no {\em bol}) to eliminate $\alpha^{fn}$ segments from the rest. We encode every class with a unique number ({\em Bol} Code) in Tab.~\ref{tbl:bol_reco_data}. 

\begin{table}[!ht]
\centering
\caption{{\em Bol} data from SR1--SR6 data sets for {\em Bol} classes\label{tbl:bol_reco_data}}
\begin{footnotesize}
\begin{tabular}{|l|r|r|r||l|r|r|r|} 
\hline
\multicolumn{1}{|c}{\textbf{\em Bol}} & \multicolumn{1}{|c}{\textbf{\em Bol}} & \multicolumn{1}{|c}{\textbf{Trg.}} & \multicolumn{1}{|c}{\textbf{Test}} & \multicolumn{1}{||c}{\textbf{\em Bol}} & \multicolumn{1}{|c}{\textbf{\em Bol}} & \multicolumn{1}{|c}{\textbf{Trg.}} & \multicolumn{1}{|c|}{\textbf{Test}} \\ 

\multicolumn{1}{|c}{\textbf{Class}} & \multicolumn{1}{|c}{\textbf{Code}} & \multicolumn{1}{|c}{\textbf{Data}} & \multicolumn{1}{|c}{\textbf{Data}} & \multicolumn{1}{||c}{\textbf{Class}} & \multicolumn{1}{|c}{\textbf{Code}} & \multicolumn{1}{|c}{\textbf{Data}} & \multicolumn{1}{|c|}{\textbf{Data}}\\ \hline \hline
a   & 1 &  139  &  34  &   ka   & 17 &  155  &  38   \\
da   & 2 &  93  &  19  &   ki   & 18 &  193  &  50   \\
dha   & 3 &  100  &  26  &   ku   & 19 &  48  &  13   \\
dhat   & 4 &  142  &  35  &   na   & 20 &  383  &  95   \\
dhi   & 5 &  55  &  16  &   ri   & 21 &  44  &  9   \\
dhin   & 6 &  324  &  81  &   ta   & 22 &  1212  &  306   \\
dhit   & 7 &  1052  &  264  &   tak   & 23 &  138  &  35   \\
ding   & 8 &  80  &  21  &   tam   & 24 &  484  &  121   \\
 e   & 9 &  142  &  35  &   tan   & 25 &  353  &  89    \\
 gadu   & 10 &  181  &  46  &   tat   & 26 &  1825  &  457    \\
 gin   & 11 &  93  &  23  &   tei   & 27 &  4060  &  1019    \\
 ha   & 12 &  447  &  112  &   tom   & 28 &  194  &  50    \\
 hat   & 13 &  171  &  42  &   tta   & 29 &  160  &  40    \\
 hi   & 14 &  154  &  37  &   ya   & 30 &  158  &  39    \\
 jag   & 15 &  22  &  8  &   yum   & 31 &  284  &  71    \\
 jham   & 16 &  32  &  11  & Stick/$\bot$ & 32 &  &   \\
 &  &  &  &  (no {\em bol}) & or 99 &  &   \\
 \hline
\end{tabular}
\end{footnotesize}
\end{table}

\subsubsection{Features of {\em Bol}s}
Since most segmented audio signals (with the exception of {\em stick-beats}) contain utterances, we use MFCC~\cite{RabinerSchafer2007} features, common for speech recognition tasks, to represent them. For each of segmented {\em bol} signal $\hat{f}^s_i(t)$ we calculate 13 dimensional MFCC features using the algorithm in~\cite{DavisMermelstein1990}. We also concatenate the dynamics features -- 13 {\em delta} and 13 {\em delta-delta} ($1^{st}$- and $2^{nd}$-order frame-to-frame difference) coefficients to get 39 features. Our choice is guided by the experimental results of~\cite{shannon2006feature} showing robustness of the extended features to background noise.

\subsubsection{GMM Training}
For recogniton of {\em bol}s from MFCC features we use GMM~\cite{reynolds2015gaussian}. GMM parameters are estimated from training data using the iterative {\em Expectation-Maximization} (EM) algorithm.

As noted in Tab.~\ref{tbl:bol_reco_data}, we have $n_b = 32$ {\em Bol} classes $C = $ $\{C_1, \cdots, C_{n_b}\}$. The GMM is trained using MFCC feature vectors of these classes. The parameters of the GMM for $C_j$ are estimated from the training data set. We train the GMM with 80\% of the {\em bol}s from the total sample of each class and save the mean, variance (diagonal covariance) and weight for each vocab class $C_j$ as the model for this class. We use $M = 15$ Gaussian components based on trials with a subset of data.

\subsubsection{Recognition Test}
We use remaining 20\% data of each class to test the model. Given the Gaussian mixture parameters for each {\em bol} class, a test vector $\mathbf{x}$ is assigned to the class $C_j$ that maximizes $p(C_j|\mathbf{x})$. That is, $p(C_j|\mathbf{x}) > p(C_i|\mathbf{x}), \forall i, 1 \leq i \leq {n_b}$. We assume that each class has equal a priori probability $p(C_j) = 1/{n_b}$. Hence, maximizing $p(C_j|\mathbf{x})$ is equivalent to maximizing $p(\mathbf{x}|C_j)$ because: $p(C_j|\mathbf{x})= p(\mathbf{x}|C_j)p(C_j) = p(\mathbf{x}|C_j)/{n_b}$.

As the occurrences of different {\em bol}s in the {\em Sollukattu}s are known, $p(C_j)$s may be estimated from these distributions provided the distribution of the {\em Sollukattu}s is known. We do not assume uniform distribution and choose to use $p(C_j) = 1/{n_b}$ for simplicity.

\subsection{Results \& Analysis}
The distribution of {\em bol}s in various classes and their partitions in training and test sets are shown in Tab.~\ref{tbl:bol_reco_data}. We have performed the {\em bol} recognition for each of SR1 through SR6 data sets and also by taking all the data sets together. In each data set we use 80\% of the data in each class for training and 20\% for testing. The results are summarized in Tab.~\ref{tbl:bol_rec}. Overall, {\bf it was possible to achieve 85.13\% accuracy in {\em bol} recognition}. 

The confusion matrix for SR1-SR6 data set is shown in Tab.~\ref{tbl:bol_conf_matrix}. From the confusion matrix we find that a group of {\em bol}s `{\em ta}', `{\em tak}', `{\em tam}', `{\em tan}' are often mis-classified among themselves. This is due to the strong similarity of their sound. 

\setlength{\tabcolsep}{2.5pt}
\begin{center}
\begin{scriptsize}
\begin{tabular}{l|rrrrrrrr|r|}
\multicolumn{1}{c}{ } &	\multicolumn{1}{c}{\textbf{\begin{rotate}{+90}ta\end{rotate}}}	&	\multicolumn{1}{c}{\textbf{\begin{rotate}{+90}tak\end{rotate}}}	&	\multicolumn{1}{c}{\textbf{\begin{rotate}{+90}tam\end{rotate}}}	&	\multicolumn{1}{c}{\textbf{\begin{rotate}{+90}tan\end{rotate}}}	&	\multicolumn{1}{c}{\textbf{\begin{rotate}{+90}tat\end{rotate}}}	&	\multicolumn{1}{c}{\textbf{\begin{rotate}{+90}tei\end{rotate}}}	&	\multicolumn{1}{c}{\textbf{\begin{rotate}{+90}tom\end{rotate}}}	&	\multicolumn{1}{c}{\textbf{\begin{rotate}{+90}tta\end{rotate}}}	&	\multicolumn{1}{c}{\textbf{\begin{rotate}{+90}Total\end{rotate}}}\\ \cline{2-10}
\textbf{ta}            & \textbf{32.4} & \alert{\bf 15.4}          & \alert{\bf 13.1}          & 0.3           & \alert{\bf 6.9}           & 0.0           & 0.3           & \alert{\bf 14.1}          & 306  \\
\textbf{tak}  & \alert{\bf 11.4}          & \textbf{68.6} & 2.9           & 0.0           & 0.0           & 0.0           & 0.0           & \alert{\bf 17.1} & 35   \\
\textbf{tam}  & \alert{\bf 9.1}           & 0.0           & \textbf{86.0} & 0.0           & 0.0           & 0.8           & 0.0           & 0.0            & 121  \\
\textbf{tan}  & \alert{\bf 15.7}          & 0.0           & \alert{\bf 46.1}          & \textbf{37.1} & 0.0           & 0.0           & 0.0           & 0.0           & 89   \\
\textbf{tat}  & 0.4           & 2.2           & 0.0           & 0.0           & \textbf{88.4} & 0.0           & 0.7           & 1.8           & 457  \\
\textbf{tei}  & 0.0           & 0.0           & 0.0           & 0.0           & 0.3           & \textbf{90.9} & 0.3           & 0.0           & 1019 \\
\textbf{tom}  & 0.0           & 0.0           & 0.0           & 0.0           & 0.0           & 4.0           & \textbf{94.0} & 0.0           & 50   \\
\textbf{tta}  & 0.0           & 5.0           & 0.0           & 0.0           & 0.0           & 0.0           & 0.0           & \textbf{95.0} & 40   \\ \cline{2-10}
\end{tabular}
\end{scriptsize}
\end{center}

\begin{table} [!ht]
\caption{Results of {\em Bol} recognition}
\label{tbl:bol_rec}
\centering
\begin{small}
\begin{tabular}{|l|r||l|r|}
\hline
\multicolumn{1}{|c}{\bf Data Set} & \multicolumn{1}{|c}{\bf Recognition} & \multicolumn{1}{||c}{\bf Data Set} & \multicolumn{1}{|c|}{\bf Recognition} \\
\multicolumn{1}{|c}{} & \multicolumn{1}{|c}{\bf Rate (\%)} & \multicolumn{1}{||c}{} & \multicolumn{1}{|c|}{\bf Rate (\%)} \\ \hline \hline
SR1 	&	85.79	&	SR4 	&	91.05	 \\	\hline
SR2 	&	87.27	&	SR5 	&	84.03	 \\	\hline
SR3 	&	94.53	&	SR6 	&	88.06	 \\	\hline
{\bf SR1--SR6} 	&	{\bf 85.13}	& \multicolumn{2}{c}{ }		 \\	\cline{1-2}
\end{tabular}
\end{small}
\end{table}

\begin{table}[!ht]
\caption{Confusion matrix for {\em Bol} recognition\label{tbl:bol_conf_matrix}}
\setlength{\tabcolsep}{2.5pt}
\centering 
\renewcommand{\arraystretch}{1.2}
\begin{scriptsize}
\begin{tabular}{l|l|r|p{1.2cm}|r||l|r|p{1.2cm}|r|}
\multicolumn{1}{c}{\multirow{27}{*}{\textbf{\begin{rotate}{+90}{Actual {\em Bol}}\end{rotate}}}}
& \multicolumn{8}{c}{\bf Predicted {\em Bol}} \\
\multicolumn{1}{c}{ } &	\multicolumn{1}{c}{\bf {\em Bol}} &	\multicolumn{1}{c}{\textbf{Self}}	&	\multicolumn{1}{c}{\textbf{Error}} &	\multicolumn{1}{c}{\textbf{Total}} &	\multicolumn{1}{c}{\bf {\em Bol}} &	\multicolumn{1}{c}{\textbf{Self}}	&	\multicolumn{1}{c}{\textbf{Error}} &	\multicolumn{1}{c}{\textbf{Total}} \\ \cline{2-9}
&	 \textbf{a}    	&	 58.8 	&	 (\alert{\bf 17.6, ta}) 	&	34	&	 \textbf{ka}   	&	 92.1 	&	 	&	38	 \\

&	 \textbf{da}   	&	 84.2 	&	 	&	19	&	 \textbf{ki}   	&	 72.0 	&	 	&	50	 \\

&	 \textbf{dha}  	&	 84.6 	&	 	&	26	&	 \textbf{ku}   	&	 92.3 	&	 	&	13	 \\

&	 \textbf{dhat} 	&	 82.9 	&	  	&	35	&	 \textbf{na}   	&	 61.1 	&	 (\alert{\bf 11.6, tam}) 	&	95	 \\

&	 \textbf{dhi}  	&	 56.3 	&	 (\alert{\bf 12.5, dhit}), (\alert{\bf 12.5, gin}) 	&	16	&	 \textbf{ri}   	&	 77.8 	&	 (\alert{\bf 11.1, dha}), (\alert{\bf 11.1, ding}) 	&	9	 \\

&	 \textbf{dhin} 	&	 87.7 	&	  	&	81	&	 \textbf{ta}   	&	 32.4 	&	 (\alert{\bf 11.4, ka}), \newline (\alert{\bf 15.4, tak}), \newline (\alert{\bf 13.1, tam}), (\alert{\bf 14.1, tta}) 	&	306	 \\

&	 \textbf{dhit} 	&	 87.9 	&	 	&	264	&	 \textbf{tak}  	&	  68.6 	&	 (\alert{\bf 11.4, ta}), \newline (\alert{\bf 17.1, tta})          	&	35	 \\

&	 \textbf{ding} 	&	 90.5 	&	 	&	21	&	 \textbf{tam}  	&	 86.0 	&	 	&	121	 \\

&	 \textbf{e}    	&	 91.4 	&	 	&	35	&	 \textbf{tan}  	&	 37.1 	&	 (\alert{\bf 15.7, ta}), \newline (\alert{\bf 46.1, tam})         	&	89	 \\

&	 \textbf{gadu} 	&	 100.0 	&	 	&	46	&	 \textbf{tat}  	&	 88.4 	&	 	&	457	 \\

&	 \textbf{gin}  	&	 82.6 	&	 	&	23	&	 \textbf{tei}  	&	 90.9 	&	 	&	1019	 \\

&	 \textbf{ha}   	&	 88.4 	&	 	&	112	&	 \textbf{tom}  	&	 94.0 	&	 	&	50	 \\

&	 \textbf{hat}  	&	 90.5 	&	 	&	42	&	 \textbf{tta}  	&	 95.0 	&	 	&	40	 \\

&	 \textbf{hi}   	&	 78.4 	&	 (\alert{\bf 16.2, e}) 	&	37	&	 \textbf{ya}   	&	 87.2 	&	 	&	39	 \\

&	 \textbf{jag}  	&	 87.5 	&	 (\alert{\bf 12.5, tat}) 	&	8	&	 \textbf{yum}  	&	 100.0 	&	 	&	71	 \\

&	 \textbf{jham} 	&	 54.5 	&	 (\alert{\bf 36.4, jag}) 	&	11	&		&		&		&		 \\

\cline{2-9}
\multicolumn{5}{c}{ } \\
\multicolumn{1}{c}{ } & \multicolumn{8}{p{7cm}}{Results for SR1--SR6 data sets (Tab.~\ref{tbl:Sollukattu_Data_Set}). For test data (Tab.~\ref{tbl:bol_reco_data}), the diagonal entries (in \%) of the confusion matrix are shown as `{\bf Self}'. Entries with 10\%+ error are shown under `{\bf Error}'. For example, for {\bf {\em Bol}} = {\bf a}, the diagonal entry is 58.8\% and it is mis-classified as {\bf ta} in 17.6\% cases. `{\bf Total}' shows the number of symbols in the class}  \\
\end{tabular}%
\end{scriptsize}
\renewcommand{\arraystretch}{1}
\end{table}

Similar cases may also be found of `{\em ta}' and `{\em tak}' mis-classified as `{\em tta}', `{\em ta}' as `{\em ka}', `{\em jham}' as `{\em jag}', and so on. To reduce mis-classifications and improve accuracy, we certainly need to improve training. For example, `{\em jham}' has only 32 training samples and ends up with 54.5\% accuracy. Interestingly, lack of training does not necessarily result in poor accuracy -- `{\em jag}' attains 87.5\% accuracy with only 22 training samples and `{\em ta}' ends up with 32.4\% in spite of having 1212 training samples. In some cases, however, completely differently sounding {\em bol}s are also mis-classified like  `{\em ri}' as `{\em dha}'. This is due to error in segmentation by silence removal because they occur side-by-side in {\em Tirmana C Sollukattu}. It was observed that the following factors influence the accuracy of {\em bol} recognition:
\begin{itemize}
\item More similar sounding {\em bol}s degrade performance. Distinctiveness of the sound of a {\em bol} helps better recognition.
\item More training samples should improve performance. 
\item The context of a {\em bol} may have significant impact on the accuracy (due to the segmentation) -- particularly the time gap with the previous and next {\em bol}s.
\end{itemize}
Using the {\em Bol} recognizer $G$ as above, we construct the Signal Signature $\Gamma(f^s)$ of the {\em Sollukattu} signal $f^s$ of $s$ and recognize $s$.

\section{{\em Sollukattu} Recognition}\label{sec:sollukattu_recognizer}
To recognize {\em Sollukattu}s we associate a {\em Signature} with a {\em Sollukattu}. Let us define a {\em Sollukattu Sequence} $\Omega(s)$ of $s \in S$ as:
\begin{eqnarray*}
\Omega(s) &=& <\omega_i\ |\ \omega_i = (\alpha_i, \beta_i), \alpha_i \in A,\\
&& \beta_i \in \mathcal{B} \cup \{\bot\}, 0 \leq i < n_{e}^s>
\end{eqnarray*}
where $A$ is the set of beat events, $i$ denotes serial position of $1$- or $\frac{1}{2}$- or $\frac{1}{4}$-beat in $s$, and $n_{e}^s$ is the number of beat ($\alpha$) events in $s$. If a beat does not have a {\em bol} ({\em stick-beat}), $\bot$ is marked for the $\beta$ event as a placeholder.

By now we know part of the $\alpha$ events ($1$-beats, $\alpha^f$ from beat detection in~\cite{mallick2018characterization}), all the $\beta$ events, and {\em Signal Signature} $\Gamma(f^{\hat{s}})$ (Sec.~\ref{sec:signal_signature}) of $f^{\hat{s}}$, where $\hat{s}$ is the placeholder for the unknown {\em Sollukattu}. We also know the time-stamps of the events, that is, $\tau(\alpha)$ and $\tau(\beta)$. So we next define {\em Signatures of Sollukattu}s.

\subsection{Signature of {\em Sollukattu}}
The {\em Signature} $\zeta(s)$ of a {\em Sollukattu} $s \in S$ is defined as:
\begin{equation}
\zeta(s) = <\beta_i\ |\ \beta_i \in \mathcal{B}, 0 \leq i < \mathcal{K}^s>
\end{equation}
by first dropping the $\alpha$ part of every $\omega$ in a {\em Sollukattu Sequence} $\Omega(s)$ and then dropping the {\em stick-beat}s ($\bot$). The signature, therefore, is a pure syntactic  representation (string over set of {\em bol}s $\mathcal{B}$) of a {\em Sollukattu} that preserves the sequencing but ignores the temporal arrangement. The length of the signature $|\zeta(s)| = \mathcal{K}^s$ denotes the number of {\em bol}s ($\beta$ events) in a bar of {\em Sollukattu} $s$. For example, for {\em Natta Sollukattu}, the {\em bol}s are as (in the notation of Sec.~\ref{sec:annotations}):

{\small	$[tei\ yum] [tat\ tat] [tei\ yum] [ta] [tei\ yum] [tat\ tat] [tei\ yum] [ta]$}

Ignoring the $\alpha$ part ($1$- or $\frac{1}{2}$-beat information), we get: 

{\small $\zeta(Natta) = tei-yum-tat-tat-tei-yum-ta-tei-yum-tat-tat-tei-yum-ta$}.

Similarly, for {\em Tirmana A}, the {\em bol}s are as: 

{\small $[ta] [hat\ ta] [jham] [ta\ ri] [ta] [B] [jham] [ta ri] [jag] [ta\ ri] [tei] [B]$}, where $[B]$ is a beat without {\em bol}. After dropping the $\alpha$, we get:

{\small $ta\ hat\ ta\ jham\ ta\ ri\ ta\ \bot\ jham\ ta ri\ jag\ ta\ ri\ tei\ \bot$}

Hence, after skipping the stick-beats, we get: 

{\small $\zeta(Tirmana\ A) = ta-hat-ta-jham-ta-ri-ta-jham-ta ri-jag-ta-ri-tei$}

All {\em Sollukattu}s, with the exception of {\em Tatta B} and {\em Tatta E}, have distinct signatures. Hence we build a dictionary $\mathcal{D}$ of {\em Sollukattu Signature}s. Once we recognize the {\em bol}s in an audio stream $f^s$ and form its {\em Signal Signature} $\Gamma(f^s)$ as the sequence of recognized {\em bol}s, we attempt to recognize the corresponding {\em Sollukattu} by matching it against the signatures in $\mathcal{D}$.

\subsection{{\em Sollukattu} Recognizer}\label{sec:sollukattu_recognition_by_sig}
Signature of the signal $\Gamma(f^{\hat{s}})$ and signatures of {\em Sollukattu}s $\zeta(s)$s are both strings defined over the same alphabet $\mathcal{B}$. However, they have different lengths. $|\Gamma(f^{\hat{s}})| = k^{\hat{s}}$ and $|\zeta(s)| = \mathcal{K}^s$, $s \in S$. Typically, the signal contains multiple cycles (bars) of the {\em Sollukattu} $\hat{s}$. Hence, often $k^{\hat{s}} \gg \mathcal{K}^s$. So every $\zeta(s)$ is repeated $\lceil k^{\hat{s}} / \mathcal{K}^s\rceil$ number of times and then $(\mathcal{K}^s-(k^{\hat{s}}\mod \mathcal{K}^s))$ symbols are truncated from the end to bring both signatures to the same length $k^{\hat{s}}$. $\zeta(s)$, so extended, is represented as $\zeta^*(s)$.

To recognize the {\em Sollukattu}, we use approximate string matching. For this we encode the {\em bol}s in the strings using the encoding scheme given in Tab.~\ref{tbl:bol_reco_data}. We then compute the {\em Best Match} between $\Gamma(f^{\hat{s}})$ and $\zeta^*(s)$, $s \in S$ using {\em Levenshtein (Edit) Distance}, $d_{Lev}$.

\subsubsection{Matching by Levenshtein (Edit) Distance}
For two strings $a$ and $b$, $d_{Lev}(a, b)$ is defined as $d_{Lev}(a, b: |a|, |b|)$ where, $d_{Lev}(a, b: i, j)$ is the distance between the first $i$ characters of $a$ and the first $j$ characters of $b$ given by:

\begin{center}
\begin{tabular}{llll}
\multicolumn{4}{l}{$d_{Lev}(a, b: i, j)$} \\
\hspace*{0.5cm} & = $max(i, j)$, if $min(i, j) = 0$ \\
                      & = $min\{d_{Lev}(a, b: i-1, j) + 1$, otherwise \\ 
                      & $\quad\quad\quad\; d_{Lev}(a, b: i, j-1) + 1$, \\
                      & $\quad\quad\quad\; d_{Lev}(a, b: i-1, j-1) + cost(a_i, b_j)\}$
\end{tabular}
\end{center}

where $cost(a_i, b_j) = 0$ if $a_i = b_j$, and $= 1$, otherwise. $d_{Lev}$ is computed assuming unit cost of insert, delete, and replace operations. We compute $d_{Lev}(\Gamma, \zeta^*(s))$, $\forall s \in S$ and find the minima. The {\em Sollukattu} of $\Gamma(f)$ is recognized as $s$ if 
$$d_{Lev}(\Gamma, \zeta(s)) = \min_{s' \in S} d_{Lev}(\Gamma, \zeta^*(s'))$$

\subsection{Results \& Analysis}
The distance matrix for SR2 data set (Tab.~\ref{tbl:Sollukattu_Data_Set}) having 40 {\em Sollukattu} audio files is shown in Tab.~\ref{tbl:edit_cmatrix}. 

\begin{table}[!ht]
\caption{Distance matrix of edit distance for SR2\label{tbl:edit_cmatrix}}
\centering
\renewcommand{\arraystretch}{1.2}
\begin{scriptsize}
\begin{tabular}{|r|r|p{4.5cm}|} 
\hline
\multicolumn{1}{|c}{\bf Test} & \multicolumn{1}{|c}{\bf Self} & \multicolumn{1}{|c|}{\bf Next min. distance or } \\ 
\multicolumn{1}{|c}{\bf {\em Sollukattu} file} & \multicolumn{1}{|c}{\bf Dist.} & \multicolumn{1}{|c|}{\bf all min. distances below the correct match} \\\hline \hline
\textbf{Joining A} 	&	\textcolor{black}{5}	& {Tattal, 17} \\ \cline{2-3}

\textbf{Joining B}	&	\textcolor{black}{0}	&	{Joining C, 11} \\ \cline{2-3}

\textbf{Joining C}	&	\textcolor{black}{1}	&	{Tatta G, 14} \\ \cline{2-3}

\textbf{Kartari} 	&	\textcolor{black}{50}	& {Tirmana B, 160} \\

\textbf{Utsanga} 	&	\textcolor{red}{\bf \em 25} & \textcolor{red}{\bf \em Nattal A, 21}; \textcolor{red}{\bf \em Tatta F, 21}; 	\textcolor{red}{\bf \em Nattal B, 22}; \newline \textcolor{red}{\bf \em Natta, 24}; \textcolor{red}{\bf \em Tatta B / E, 25};  \textcolor{red}{\bf \em Joining A, 25}\\ 

\textbf{Mandi 1}	&	\textcolor{black}{70}	&	{Natta, 292} \\

\textbf{Sarikkal 1}	&	\textcolor{black}{64}	&	{Natta, 316}; {Tatta F, 316}	\\

\textbf{Sarikkal 2} 	&	\textcolor{black}{29}	&	{Tirmana B, 72}	\\ \cline{2-3}

\textbf{Kuditta Mettu 1} 	&	\textcolor{black}{0}	&	{Sarika, 32}	\\

\textbf{Kuditta Mettu 2} 	&	\textcolor{black}{2}	& {Sarika, 18} \\

\textbf{Kuditta Mettu 3}	&	\textcolor{black}{0}	&	{Sarika, 64} \\ \cline{2-3}

\textbf{Kuditta Nattal A1} 	&	\textcolor{black}{7}	&	{Nattal B, 21}	\\

\textbf{Kuditta Nattal A2} 	&	\textcolor{black}{29}	&	{Nattal B, 35}	\\ \cline{2-3}

\textbf{Kuditta Nattal B1}	&	\textcolor{black}{5} 	&	{Nattal A, 21}	\\

\textbf{Kuditta Nattal B2}	&	\textcolor{black}{2} 	&	{Nattal A, 18}	\\ \cline{2-3}

\textbf{Kuditta Tattal 1} 	&	\textcolor{black}{3}	&	{Paikkal, 120}; {TTD, 120}	\\

\textbf{Kuditta Tattal 2} 	&	\textcolor{black}{4}	&	{TTD, 221}	\\

\textbf{Kuditta Tattal 3} 	&	\textcolor{black}{0}	&	{Paikkal, 80}; {Pakka, 80}; {TTD, 80} \\

\textbf{Kuditta Tattal 4}	&	\textcolor{black}{2}	&	{Paikkal, 41}; {TTD, 41} \\ \cline{2-3}

\textbf{Natta 1}	&	\textcolor{black}{2}	& {Tatta F, 23} \\

\textbf{Natta 2}	&	\textcolor{red}{\bf \em 33}	&	\textcolor{red}{\bf \em Tatta F, 29} \\

\textbf{Natta 3}	&	\textcolor{black}{11}	&	{Tatta F, 99} \\ \cline{2-3}

\textbf{Paikkal 1} 	&	\textcolor{black}{27}	&	{Tatta F, 103} \\ \cline{2-3}

\textbf{Pakka 1} 	&	\textcolor{black}{4}	&	{Tatta F, 61} \\

\textbf{Pakka 2}	&	\textcolor{black}{4}	&	{Tatta F, 120} \\

\textbf{Pakka 43}	&	\textcolor{black}{4}	&	{Tatta F, 32}; {Tatta G, 32}; {TTD, 32} \\ \cline{2-3}

\textbf{Sarika 1} 	&	\textcolor{black}{5}	&	{Nattal A, 63} \\

\textbf{Sarika 2}	&	\textcolor{black}{7}	&   {Nattal A, 13}	\\ \cline{2-3}

\textbf{Tatta A}	&	\textcolor{black}{26}	&	{Tatta C, 27} \\ \cline{2-3}

\textbf{Tatta B}	&	\textcolor{black}{0}	&   {Tatta F, 9} \\ \cline{2-3}

\textbf{Tatta C}	&	\textcolor{black}{2}	&	{Tatta A, 12} \\ \cline{2-3}

\textbf{Tatta D} 	&	\textcolor{black}{0}	& {Tatta B / E, 11} \\ \cline{2-3}

\textbf{Tatta E} 	&	\textcolor{black}{5}	&	{Tatta A, 20}; {Tatta F, 20} \\ \cline{2-3}

\textbf{Tatta F} 	&	\textcolor{black}{1}	&	{Tatta B / E, 22} \\ \cline{2-3}

\textbf{Tatta G}	&	\textcolor{black}{1}	&	{KUMS, 19} \\ \cline{2-3}

\textbf{Tei Tei Dhatta 1} 	&	\textcolor{black}{33}	&	{Pakka, 58} \\

\textbf{Tei Tei Dhatta 2}	& \textcolor{black}{20}	& {Pakka, 36} \\ \cline{2-3}

\textbf{Tirmana A} 	&	\textcolor{black}{25}	&	{Tattal, 44}; {TTD, 44} \\ \cline{2-3}

\textbf{Tirmana B}	&	\textcolor{black}{24}	&	{Tirmana C, 135} \\ \cline{2-3}

\textbf{Tirmana C}	&	\textcolor{black}{111} & {Tirmana B, 201} \\ \hline
\multicolumn{3}{l}{ } \\

\multicolumn{3}{p{7.5cm}}{$\bullet$ KUMS is Kartati Utsanga Mandi Sarikkal, Mettu is Kuditta Mettu, Nattal is Kuditta Nattal, Tattal is Kuditta Tattal and Tei Tei Dhatta is TTD} \\

\multicolumn{3}{p{7.5cm}}{$\bullet$ Self Distance is the distance with the correct entry for the input {\em Sollukattu} in the dictionary. If it is the minimum (correct match), we show the next minimum for an estimate of discrimination of edit distance. If it is not the minimum (wrong match), we show all distances that are smaller than it.} \\

\multicolumn{3}{p{7.5cm}}{$\bullet$ Some {\em Sollukattu}s have multiple recorded files as shown with serial numbers. 40 files for 23 {\em Sollukattu}s in SR2} \\

\multicolumn{3}{p{7.5cm}}{$\bullet$ Two {\em Sollukattu}s (shown in red, bold) are wrongly classified. Hence the accuracy is $(40-2)/40 = 95\%$ (Tab.~\ref{tbl:sol_rec})} \\

\multicolumn{3}{p{7.5cm}}{$\bullet$ {\em Tatta B} and {\em Tatta E} have the same signature (differ only in {\em stick-beats})} 
\end{tabular}
\end{scriptsize}
\renewcommand{\arraystretch}{1}
\end{table}

We see that recognition fails for two {\em Sollukattu}s:
\begin{itemize}
\item {\tt Utsanga\_HB2.wav} ({\em KUMS}) is mis-classified as {\em Kuditta Nattal A}. It may noted (Tab.~\ref{tbl:sollukattu_error}(a)) that the key {\em bol} `{\em tan}' of {\em KUMS} is repeatedly mis-classified, often as `{\em tam}'. The other key {\em bol} `{\em gadu}' is often missed. In contrast, `{\em tei}' is getting recognized which does not exist in this {\em Sollukattu}. This results in a stronger similarity with and classification to {\em Kuditta Nattal A}.

\item {\tt Natta\_35678\_HB2.wav} ({\em Natta}) is mis-classified as {\em Tatta F}. The {\em bol} `{\em yum}' of {\em Natta} is totally missing making it very close to the signature of {\em Tatta F} (Tab.~\ref{tbl:sollukattu_error}(b)).
\end{itemize}

\begin{table}[!ht]
\caption{Error cases in {\em Sollukattu} recognition\label{tbl:sollukattu_error}}
\begin{center}
\begin{scriptsize}
\begin{tabular}{|lllllllllllllll|} \multicolumn{15}{c}{$\Gamma($\tt Utsanga\_HB2.wav$)$} \\ \hline
hi & tam & gadu & tat & tat & na & tam & tat & tat & tei & ta & tam & tat & tei & $\cdots$ \\
14 & 24 & 10 & 26 & 26 & 20 & 24 & 26 & 26 & 27 & 22 & 24 & 26 & 27 & $\cdots$ \\ \hline
\multicolumn{15}{c}{ } \\ 
\multicolumn{15}{c}{$\zeta($\em KUMS$)$: [tan gadu] [tat tat] [dhin na] [tan gadu] [tat tat] [dhin na]} \\ \hline
tan & gadu 	& tat 	& tat 	& dhin 	& na  & tan & gadu 	& tat 	& tat 	& dhin 	& \multicolumn{1}{l|}{na} & tan & gadu & $\cdots$ 	\\
25 	& 10	& 26 	& 26 	& 06 	& 20 & 25 	& 10 	& 26 	& 26 	& 06 	& \multicolumn{1}{l|}{20} & 25 	& 10 & $\cdots$ \\ \hline
\multicolumn{15}{c}{ } \\ 
\multicolumn{15}{c}{$\zeta($\em Kuditta Nattal A$)$: [tat] [tei] [tam] [B] [dhit] [tei] [tam] [B]} \\ \hline
tat & tei & tam & dhit & tei & \multicolumn{1}{l|}{tam} & tat & tei & tam & dhit & tei & \multicolumn{1}{l|}{tam} & tat & tei & $\cdots$\\
26 & 27 & 24 & 07 & 27 & \multicolumn{1}{l|}{24} & 26 & 27 & 24 & 07 & 27 & \multicolumn{1}{l|}{24} & 26 & 27 & $\cdots$\\ \hline
\multicolumn{15}{c}{ } \\
\multicolumn{15}{c}{(a)} \\
\multicolumn{15}{c}{ } \\
\multicolumn{15}{c}{$\Gamma($\tt Natta\_35678\_HB2.wav$)$} \\ \hline
tei & tat & tat & tei & ta & tei & tat & tei & ta & tei & tat & tat & tei & ta & $\cdots$ \\
27 & 26 & 26 & 27 & 22 & 27 & 26 & 27 & 22 & 27 & 26 & 26 & 27 & 22 & $\cdots$ \\ \hline
\multicolumn{15}{c}{ } \\ 
\multicolumn{15}{c}{$\zeta($\em Natta$)$: [tei yum] [tat tat] [tei yum] [ta] [tei yum] [tat tat] [tei yum] [ta]} \\ \hline
tei & yum 	& tat 	& tat 	& tei 	& yum 	& ta & tei 	& yum 	& tat 	& tat 	& tei 	& yum 	& \multicolumn{1}{l|}{ta} & $\cdots$\\
27 	& 31 	& 26 	& 26 	& 27 	& 31 	& 22 & 27 	& 31 	& 26 	& 26 	& 27 	& 31 	& \multicolumn{1}{l|}{22} & $\cdots$\\ \hline
\multicolumn{15}{c}{ } \\ 
\multicolumn{15}{c}{$\zeta($\em Tatta F$)$: [tei] [tei] [tat] [tat] [tei] [tei] [tam] [B]} \\ \hline
tei & tei & tat & tat & tei & tei & \multicolumn{1}{l|}{tam} & tei & tei & tat & tat & tei & tei & \multicolumn{1}{l|}{tam} & $\cdots$\\
27 & 27 & 26 & 26 & 27 & 27 & \multicolumn{1}{l|}{24} & 27 & 27 & 26 & 26 & 27 & 27 & \multicolumn{1}{l|}{24} & $\cdots$\\ \hline
\multicolumn{15}{c}{ } \\
\multicolumn{15}{c}{(b)} \\
\end{tabular}
\end{scriptsize}
\end{center}
\end{table}

Of the total 162 {\em Sollukattu} recordings, 8 were partially corrupted and could not be used. Of the remaining 154 files, 7 were mis-classified. So {\bf we achieve 95.45\% accuracy in {\em Sollukattu} recognition (Tab.~\ref{tbl:sol_rec})}. Next we estimate the tempo period.
\begin{table*} [!ht]
\caption{Results of {\em Sollukattu} recognition\label{tbl:sol_rec}}
\centering
\begin{scriptsize}
\begin{tabular}{|l|r|r|r|l|l|}
\hline
\multicolumn{1}{|c}{\bf Data} & \multicolumn{1}{|c}{\bf No. of}  & \multicolumn{1}{|c}{\bf No. Correctly}  & \multicolumn{1}{|c}{\bf \% Rate of} & \multicolumn{1}{|c}{\bf Mis-classification} & \multicolumn{1}{|c|}{\bf Remarks} \\ 
\multicolumn{1}{|c}{\bf Set} & \multicolumn{1}{|c}{\bf Audio Files}  & \multicolumn{1}{|c}{\bf Recognized}  & \multicolumn{1}{|c}{\bf Recognition} & \multicolumn{1}{|c}{\bf File: Actual Sollukattu $\rightarrow$  Predicted Sollukattu} & \multicolumn{1}{|c|}{\bf } \\ \hline \hline
SR1	&	30	&	29	&	96.67	&	Tirmana\_1\_HB1.wav: Tirmana A $\rightarrow$ Joining A	&		\\ \hline
SR2	&	40	&	38	&	95.00	&	Natta\_35678\_HB2.wav: Natta  $\rightarrow$ Tatta F	&	Edit distance matrix shown in Tab.~\ref{tbl:edit_cmatrix}	\\ 
	&		&		&		&	Utsanga\_HB2.wav: KUMS  $\rightarrow$ Kuditta Nattal A	&		\\ \hline
SR3	&	20	&	19	&	95.00	&	Tatta\_4\_HB4.wav: Tatta C  $\rightarrow$ Tatta D	&	3 files were partially corrupted \& skipped: 	\\ 
	&		&		&		&		&	Joining A, Kuditta Nattal A, KUMS	\\ \hline
SR4	&	22	&	21	&	95.45	&	Tatta\_4\_HB4.wav: Tatta C  $\rightarrow$ Tatta D	&	1 file was partially corrupted \& skipped: 	\\ 
	&		&		&		&		&	Tatta B	\\ \hline
SR5	&	22	&	20	&	90.91	&	Tatta\_12\_MD1.wav: Tatta A  $\rightarrow$ Tatta D	&	1 file was partially corrupted \& skipped: 	\\ 
	&		&		&		&	Tatta\_4\_MD1.wav: Tatta C  $\rightarrow$ Tatta A	&	Joining C	\\ \hline
SR6	&	20	&	20	&	100.00	&		&	3 files were partially corrupted \& skipped: 	\\
	&		&		&		&		&	Kuditta Tattal, KUMS, Tatta A	\\ \hline\hline
{\bf Total}	&	{\bf 154}	&	{\bf 147}	&	{\bf 95.45}	&		&		\\ \hline
\multicolumn{6}{l}{ } \\
\multicolumn{6}{l}{$\bullet$ Data Sets from Tab.~\ref{tbl:Sollukattu_Data_Set}. KUMS is Kartati Utsanga Mandi Sarikkal}
\end{tabular}
\end{scriptsize}
\end{table*}

\section{Tempo Period Estimation}\label{sec:tempo_period_estimation}
In an earlier paper~\cite{mallick2018characterization} we reported the detection of $1$-beats with their time-stamps. 
This approach can be used to compute the tempo period from the difference of time-stamps of two consecutive $1$-beats. This difference should be identical for two consecutive $1$-beats and be the same as $T^s$. An estimator may also be designed based on the time-stamps of detected {\em bol}s. However, these strategies do not work well due to human errors in creating the signal and due to limitations of the detection algorithms. We outline the issues below:
\begin{enumerate}
\item The difference of time-stamp of two consecutive $1$-beats vary (substantially at times) due to the human error in beating the stick or uttering the {\em bol}s or both. Situation worsens when the {\em Sollukattu} has $\frac{1}{2}$- and $\frac{1}{4}$-beats.
\item The segmentation of $f^s(t)$ by silence (Sec.~\ref{sec:audio_segmentation}) may have some errors. This propagates to the estimated time-stamp of the detected {\em bol}.
\item Many {\em Sollukattu}s have $\frac{1}{2}$-beats, some even have $\frac{1}{4}$-beats. Due to the error in {\em bol} recognition, at times it may not be possible to correctly identify if a slice (and its time-stamp) corresponds to a $1$-beat or a $\frac{1}{2}$-beat. This may cause errors in time gaps.
\end{enumerate}
We explore two approaches for estimation of tempo period:
\begin{itemize}
\item Estimation from the audio signal $f^s$ using {\em Resonating / Comb Filter}
\begin{itemize}
\item This operates at the low level, working directly with the signal 
\end{itemize}
\item Estimation using {\em Longest Common Sub-string (LCS)} between {\em Signal signature} and {\em Sollukattu Signature}
\begin{itemize}
\item This operates at the high level, exploiting the structural information extracted so far
\end{itemize}
\end{itemize}

\subsection{Estimation using Comb Filter}
A comb filter is often used for tempo estimation and beat tracking in music signal processing (\cite{scheirer1998tempo}). If a piece of music can be characterized as consisting of musical events which are often {\em on the beat}, then we may expect that signal processing methods such as comb filtering and auto-correlation to succeed in locating the beats. 

To estimate the tempo period, we use the method in~\cite{scheirer1998tempo}. 
\begin{enumerate}
\item \textbf{Frequency Filter-bank}: First the audio signal $f^s$ is passed through a bank of 3 filters corresponding to 3 typical bands; namely, vocal (0-900Hz), instrumental beating (900-2600Hz) and harmonics (2600-22100Hz). Output of each filter, in time domain, is processed through the following steps.

\item \textbf{Envelop Extractor}: The signal $f^s$ has a range of frequencies in every band. However, we are interested only in the overall trend (the slow periodicity of the signal devoid of the fine changes at every frequency). And we expect this trend to be similar in every band. So we need to compute the envelop of the signal where only sudden changes in the signal can strongly manifest. Naturally, we need to filter the frequency-banded signals using low-pass filters. Hence, we first full-wave rectify the signals to reduce high-frequency content and restrict to the positive half of the envelope. We then convolve each signal by the right half of a {\em Hanning Window} for low-pass filtering. Much of this computation is performed in the frequency domain for ease of implementation and computational efficiency.

\item \textbf{Differentiator}: The envelop signal is next differentiated in time domain to manifest sudden changes in its amplitude.

\item \textbf{Half-wave Rectifier}: The differentiated signal is half-wave rectified to enhance the changes. Now the signal resembles a sequence of (imperfect) impulses. The temporal periodicity in these impulses indicate the tempo period. We intend to estimate that by combing.

\item \textbf{Resonant Filter-bank}: We construct a set of equi-spaced impulse trains having adjustable periodicity (spacing) for impulses. We expect that the periodicity of one of these impulse trains will match the periodicity of our sequence of impulses. So if we convolve the impulse train with our sequence, the resulting energy will maximize ({\em resonate}) when their periods match. We perform this in frequency domain with periodicity varying from 33 bpm (bpm$_{min}$ = 33 $\approx$ 1.8 sec.) to 75 bpm (bpm$_{max}$ = 75 $\approx$ 0.8 sec.) in unit steps. We convolve to compute the energy in each band. The bpm $p$ corresponding to maximum sum of energy is taken to be the fundamental bpm of the audio signal. The tempo period is then computed as 60/$p$ sec.
\end{enumerate}

The results of tempo period estimation by this methods are shown in Tab.~\ref{tbl:tempo_period_results}. We analyze in Sec.~\ref{sec:tempo_period_result}.

\subsection{Estimation using LCS between Signatures}
After recognizing the {\em Sollukattu} $s$ for test signal $f^s$, we know the following:
\begin{itemize}
\item The {\em Signal Signature} $\Gamma(f^s)$ has detected sequence of {\em bol}s with time-stamps. Some {\em bol}s in this sequence may be wrong, time-stamps may be erroneous, and {\em bol}s may be at $1$-beat or $\frac{1}{2}$-beat positions (and we do not know which is at $1$-beat and which is at $\frac{1}{2}$-beat). So we cannot compute the tempo period ($1$-beat to $1$-beat gap) directly from such a signature. 
\item The {\em Sollukattu Signature} $\zeta(s)$ having correct sequence of {\em bol}s. We also know if these {\em bol}s are at $1$- or $\frac{1}{2}$-beats positions. However, we do not know their time-stamps.
\end{itemize}

If the recognition of {\em bol}s in $\Gamma(f^s)$ were all correct, we could just match it up with $\zeta^*(s)$ (Sec.~\ref{sec:sollukattu_recognition_by_sig}) symbol by symbol to know the {\em bol}s at $1$-beat and then use their time-stamps to get the tempo period. This is not possible because there will be wrong {\em bol}s in $\Gamma(f^s)$. However, if we assume that most {\em bol}s are correctly recognized in $\Gamma(f^s)$ (which is often the case), then we can expect long sub-strings of {\em bol}s that are correct and match them against $\zeta(s)$. The longer the sub-string, better will be the quality of the match. Hence we look for longest common sub-string $\mathcal{L}(\Gamma,\zeta)$ between $\Gamma(f^s)$ and $\zeta(s)$. It is expected that such a sequence will be long enough to contain multiple $1$-beats ($\alpha^f_i$, $0 \leq i < |\mathcal{L}(\Gamma,\zeta)|$) that can now be correctly known for their time-stamps. Thus one can have a number of estimates ($T_{est}^i = \tau(\alpha^f_{i+1})-\tau(\alpha^f_{i})$, $0 < i \leq |\mathcal{L}(\Gamma,\zeta)|$) for the tempo period and choose their median as a robust estimator. Formally:

\subsubsection{Longest Common Sub-strings (LCS)}
Let $a$ be a string of length $|a| = n$. $a_{i..j}$, $i \leq j$ is a sub-string of $a$ containing $i^{th}$ to $j^{th}$ symbols. For two strings $a$ ($|a| = n$) and $b$ ($|b| = m$) over the same alphabet, the length of the longest common suffix for all pairs of prefixes of the strings are defined by the following recursion:

\begin{center}
\begin{tabular}{llll}
\multicolumn{4}{l}{$LCS_{Suff}(a_{1..n}, b_{1..m})$} \\
\hspace*{0.5cm} & = $LCS_{Suff}(a_{1..n-1}, b_{1..m-1}) + 1$, if $a_{n} = b_{m}$ \\
                      & = 0, otherwise \\
\end{tabular}
\end{center}

$LCS_{Suff}$ can be computed efficiently using dynamic programming. The longest suffix strings may be constructed by tracing back on the updates of the DP tableau. The length of the longest common sub-strings (LCS) of $a$ and $b$ is the maximum of the lengths of the longest common suffixes:
\[
 LCS(a_{1..n}, b_{1..m}) = \max_{\substack{
 1\leq i \leq n, 1\leq j \leq m
 }} LCS_{Suff}(a_{1\cdots i}, b_{1\cdots j})
\]
There may be multiple suffixes having this maximum value. All of them are longest common sub-strings of $a$ and $b$. 

Consider {\em Joining B}. Given $\zeta(Joining\ B)$:

\begin{center}
\begin{footnotesize}
\begin{tabular}{|llllllllllll|} 
\hline
dhit	& dhit 	& tei 	& dhit 	& dhit 	& tei & dhit & dhit & tei & dhit & dhit & tei  \\
7 		& 7 	& 27 	& 7 	& 7 	& 27  & 7 & 7 & 27 & 7 & 7 & 27 \\
B & HB  & B & B & HB & B  & B & HB & B & B & HB & B \\ \hline
\end{tabular}
\end{footnotesize}
\end{center}

and $\Gamma(f^{Joining\ B})$ for a sample audio: 

\begin{footnotesize}
[7--7--27--7--7--27--7--7--27--7--7--27--7--7--27--7--7--27--7--7--27--7--7--27]
\end{footnotesize}

the LCS and time-stamps are obtained as:

\begin{center}
\begin{footnotesize}
\begin{tabular}{|l|r|r|r|r|r|r|} 
\hline
LCS, $\mathcal{L}(\Gamma,\zeta)$      &   07 &   07 &   27 &   07 &   07 &   27 \\
$\alpha$ &    B &   HB &    B &    B &   HB &    B  \\
$\tau(\alpha)$ in sec.    & 1.04 & 1.97 & 2.90 & 4.58 & 5.38 & 6.19  \\
$T_{est}$ in sec. & 1.86 &      & 1.68 & 1.61 &      & 1.53 \\ \hline \hline
LCS, $\mathcal{L}(\Gamma,\zeta)$  & 07 &   07 &   27 &    07 &    07 & 27 \\
$\alpha$ & B &   HB &    B &     B &    HB & B \\
$\tau(\alpha)$ in sec. & 7.72 & 8.46 & 9.23 & 10.62 & 11.37 & 12.11 \\
$T_{est}$ in sec. & 1.51 &      & 1.39 &  1.49 &       &   \\ \hline
\end{tabular}
\end{footnotesize}
\end{center}

Hence, the estimated tempo period is the median of row $T_{est}$, that is, 1.53 sec. The annotated time for this is 1.52 sec.

\subsection{Results \& Analysis}\label{sec:tempo_period_result}
\subsubsection{By Comb filter}
The estimation of tempo period by Comb filter has been tested with SR1 data set (Tab.~\ref{tbl:Sollukattu_Data_Set}). The result is given in Tab.~\ref{tbl:tempo_period_results}. All tempo periods have been accurately estimated with the sole exception of {\em Joining B Sollukattu}. So {\bf we could achieve 96.67\% accuracy in estimation of tempo period by Comb filter}.

\begin{table}[!ht]
\caption{Results of tempo estimation from {\em Bol}s\label{tbl:tempo_period_results}}
\centering
\begin{scriptsize}
\begin{tabular}{|l|r||r|r||r|r|} 
\cline{3-6}
\multicolumn{2}{c}{ } & \multicolumn{4}{|c|}{\bf Estimation Methods} \\ \cline{3-6}
\multicolumn{2}{c}{ } & \multicolumn{2}{|c}{\bf Comb Filter} & \multicolumn{2}{||c|}{\bf LCS} \\ \hline
\multicolumn{1}{|c}{\textbf{\em Sollukattu}}	&	\multicolumn{1}{|c}{\textbf{Actual}}	&	\multicolumn{1}{||c}{\textbf{Est.}}	&	\multicolumn{1}{|c}{\textbf{Abs.}} &	\multicolumn{1}{||c}{\textbf{Est.}}	&	\multicolumn{1}{|c|}{\textbf{Abs.}}	\\

\multicolumn{1}{|c}{\textbf{ Name}}	&	\multicolumn{1}{|c}{\textbf{Tempo}}	&	\multicolumn{1}{||c}{\textbf{Tempo}}	&	\multicolumn{1}{|c}{\textbf{Error}} &	\multicolumn{1}{||c}{\textbf{Tempo}}	&	\multicolumn{1}{|c|}{\textbf{Error}} \\ \hline \hline
Joining A	&	1.18	&	1.15	&	0.03	&	1.22	&	0.04	\\ \hline
Joining B	&	1.52	&	0.80	&	\alert{\bf 0.72}	&	1.53	&	0.01	\\ \hline
Joining C	&	1.17	&	1.15	&	0.02	&	1.17	&	0.00	\\ \hline
Kartari Utsanga	&	1.07	&	1.02	&	0.05	&	1.11	&	0.04	\\ \cline{2-6}
Mandi Sarikkal	&	1.00	&	1.09	&	0.09	&	1.05	&	0.05	\\ \hline
Kuditta Mettu 1	&	1.16	&	1.15	&	0.01	&	1.16	&	0.00	\\ \cline{2-6}
Kuditta Mettu 2	&	1.16	&	1.07	&	0.09	&	1.08	&	0.08	\\ \hline
Kuditta Nattal A	&	0.99	&	0.98	&	0.01	&	1.05	&	0.06	\\ \hline
Kuditta Nattal B	&	1.30	&	1.30	&	0.00	&	1.31	&	0.01	\\ \hline
Kuditta Tattal 1	&	1.21	&	1.20	&	0.01	&	1.22	&	0.01	\\ \cline{2-6}
Kuditta Tattal 2	&	1.21	&	1.15	&	0.06	&	1.13	&	0.08	\\ \cline{2-6}
Kuditta Tattal 3	&	1.21	&	1.09	&	0.12	&	1.10	&	0.11	\\ \cline{2-6}
Kuditta Tattal 4	&	1.21	&	1.18	&	0.03	&	1.15	&	0.06	\\ \hline
Natta 1	&	1.39	&	1.40	&	0.01	&	1.38	&	0.01	\\ \cline{2-6}
Natta 2	&	1.39	&	1.36	&	0.03	&	1.36	&	0.03	\\ \hline
Paikkal	&	1.58	&	1.58	&	0.00	&	1.55	&	0.03	\\ \hline
Pakka 1	&	1.21	&	1.20	&	0.01	&	1.21	&	0.00	\\ \cline{2-6}
Pakka 2	&	1.21	&	1.15	&	0.06	&	1.14	&	0.07	\\ \hline
Sarika	&	0.93	&	0.92	&	0.01	&	0.90	&	0.03	\\ \hline
Tatta A	&	1.51	&	1.50	&	0.01	&	1.52	&	0.01	\\ \hline
Tatta B	&	1.36	&	1.33	&	0.03	&	1.35	&	0.01	\\ \hline
Tatta C	&	1.56	&	1.58	&	0.02	&	1.55	&	0.01	\\ \hline
Tatta D	&	1.35	&	1.36	&	0.01	&	1.34	&	0.01	\\ \hline
Tatta E	&	1.17	&	1.18	&	0.01	&	1.20	&	0.03	\\ \hline
Tatta F	&	1.21	&	1.20	&	0.01	&	1.25	&	0.04	\\ \hline
Tatta G	&	1.41	&	1.30	&	0.11	&	1.32	&	0.09	\\ \hline
Tei Tei Dhatta	&	1.41	&	1.40	&	0.01	&	1.41	&	0.00	\\ \hline
Tirmana A	&	1.23	&	1.22	&	0.01	&	\alert{\bf fail}	&	\alert{\bf fail}	\\ \hline
Tirmana B	&	1.22	&	1.18	&	0.04	&	1.21	&	0.01	\\ \hline
Tirmana C	&	1.46	&	1.36	&	0.10	&	1.35	&	0.11	\\ \hline
\multicolumn{6}{c}{ } \\
\multicolumn{6}{p{6.25cm}}{$\bullet$ Results for SR1 data set (Tab.~\ref{tbl:Sollukattu_Data_Set}). Multiple samples from a {\em Sollukattu} are serially numbered} \\
\multicolumn{6}{p{6.25cm}}{$\bullet$ Tempo period for 29 out of 30 sample signals are correctly estimated by each method. Errors in estimation are highlighted} 
\end{tabular}
\end{scriptsize}
\end{table}

\begin{figure}
\centering
\includegraphics[width=8cm]{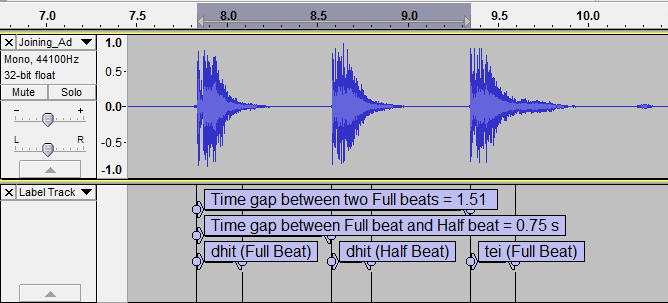}
\caption{Signal of {\em Joining B Sollukattu}\label{fig:tempo_error_joining_2}}
\end{figure}

The tempo period of {\em Joining B} has been estimated as 0.72 sec. while it actually is 1.52 sec., that is, almost the double. Checking the signal (Fig.~\ref{fig:tempo_error_joining_2}) we find that in this case the $\frac{1}{2}$-beats have same energy as the $1$-beat. Hence, there are equal peaks at $\frac{1}{2}$-beats as well and the fundamental bpm has been computed based on the number of $\frac{1}{2}$- and $1$-beats instead of just the $1$-beats. Such errors, however, may be easily corrected once the {\em Sollukattu} has been recognized and $\frac{1}{2}$-beats are known.

\subsubsection{By LCS}
The estimation of tempo period by LCS has also been tested with SR1 data set (Tab.~\ref{tbl:Sollukattu_Data_Set}). The result is given in Tab.~\ref{tbl:tempo_period_results}. We find that this method is as accurate as the Comb filter based method and {\bf we achieve 96.67\% accuracy in estimation of tempo period by LCS}.

This algorithm, however, fails for {\em Tirmana A}. Due to error in {\em bol} recognition, the LCS in this case contains only one $1$-beat. Hence the tempo period cannot be computed.

We use the tempo period detected by LCS method for beat marking. Comb filter based tempo period is used when the LCS based method fails. 

\section{Beat Marking}\label{sec:beat_marking}
Now we are ready to mark the beats ($\alpha$ events) on the audio signal $f^s(t)$. This would involve the following:
\begin{enumerate}
\item Mark time-stamps on $f^s(t)$ that are beats.
\item Annotate every such marking as a $1$-beat ($\alpha^{fb}$), $\frac{1}{2}$-beat ($\alpha^{hb}$), or stick-beat ($\alpha^{fn}$). Note that $\frac{1}{4}$-beats ($\alpha^{qb}$) are not considered.
\item For a $1$- ($\alpha^{fb}$) or a $\frac{1}{2}$-beat ($\alpha^{hb}$), annotate the {\em bol} symbol.
\end{enumerate}

With this it would be possible to automatically generate audio annotations as shown in Fig.~\ref{fig:Tatta_C_Annotations}.

For this task we use the following information as extracted:
\begin{enumerate}
\item {\em Detected Beats}: Array ($DB$) of $1$-beats extracted using onset (as in~\cite{mallick2018characterization}). Each beat has a time-stamp. 
\item {\em Signal Signature}: Array ($SS$) of {\em bol}s extracted through {\em bol} recognition (Sec.~\ref{sec:signal_signature},~\ref{sec:bol_segmentation}). Each {\em bol} has an associated time-interval (from the non-silent slice).
\item Estimated tempo period $T$ (Sec.~\ref{sec:tempo_period_estimation}).
\end{enumerate} 

In addition, we classify non-silent slices $\hat{f}^s_i(t)$, $0 \leq i < k^s -1$  (Eqn.~\ref{eqn:non_silent_slices}) of the {\em bol}s as having $high$ or $low$ energy. We calculate energy of every slice as in Sec.~\ref{sec:silence_removal} and cluster the values by $k$-means clustering with $k = 2$. The energy class $\epsilon$ ($high$ or $low$) is then marked on {\em Signal Signature} array $SS$.
 
To mark and annotate $f^s(t)$, we note the following structural properties of a {\em Sollukattu}:
\begin{enumerate}
\item The time gap between consecutive $1$-beats should approximately match the tempo period $T$. 
\begin{itemize}
\item We use a wider period $wide(T) = [T-0.25, T+0.4]$ for a beat-to-beat gap.
\end{itemize}
\item A {\em bol} always occurs with a beat. Hence its offset from the previous $1$-beat should 
\begin{itemize}
\item approximately match the tempo period ($wide(T)$) if it occurs with a $1$-beat, or
\item be less than the tempo period if it occurs with a $\frac{1}{2}$-beat
\end{itemize}
\item A $1$-beat with {\em bol} has higher energy than a stick-beat.
\begin{itemize}
\item This may help filter wrongly detected {\em bol}s (from stick-beats) from being marked as a $1$-beat.
\end{itemize}
\item There should be a $1$-beat at the approximate periodicity of the tempo period.
\begin{itemize}
\item If there is a long gap between $1$-beats, say, more than $long\_gap(T) = 2*T-0.25$; a beat must have been missed and should be assumed.
\end{itemize}
\end{enumerate} 

Using $DB$, $SS$, and $T$ as input, we compute the beat marking information as an array $MB$ in Algorithm~\ref{algo_bm}. $MB$ carries the information of every beat type ($1$-, $\frac{1}{2}$- or stick), the associated time interval and the {\em bol}, if any. The algorithm follows the structural properties stated above to compute $MB$.

\renewcommand{\baselinestretch}{1} 
\begin{algorithm}
\caption{: Beat Marking}\label{algo_bm}
 \begin{footnotesize}
 \begin{algorithmic}[1]
 \State {\bf Inputs:}
 \State $DB$ = Array of time-stamps of $1$-beats detected by onset 
 \State $SS$ = Signal Signature (Sec.~\ref{sec:signal_signature}) as an array. Each element is a quadruple comprising [{\em Bol} $\beta$, Energy $\epsilon(\beta)$, Start time $\tau_s(\beta)$, End time $\tau_e(\beta)$]. Energy has values $high$ or $low$.
	\State $T$ = Tempo Period as estimated in Sec.~\ref{sec:tempo_period_estimation}
	\State {\bf Output:}
	\State $MB$ = Sequence of marked beats as an array. Each element is a quadruple comprising [{\em Bol} $\beta$, Start time $\tau_s(\beta)$, End time $\tau_e(\beta)$, Beat Info]. Beat Info is an audio event -- one of $\alpha^{fb}$, $\alpha^{hb}$, $\alpha^{fn}$.
	\State {\bf Steps:}
 \State	/* Computation of $OB$ = Binary array of overlapped beats. $OB(i)$ is set to $true$ if there exists a beat $b$ with time $\tau(b) \in DB$ such that $\tau(b)$ lies within $[SS(i).\tau_s(\beta),SS(i).\tau_e(\beta)]$. That is, the $i^{th}$ {\em bol} $SS(i).\beta$ overlaps with a beat from onset. Otherwise, it is set to $false$. */
 \State $p \leftarrow 1$; 
 \While {$p < length(SS)$} /* Initialize $OB$ to $false$ */
 \State $OB \leftarrow false$;
 \EndWhile
 \State $p \leftarrow 1$; $q \leftarrow 1$;
 \While {$((p < length(SS))$ \& $(q < length(DB)))$}
 \If {$(DB(p) < SS(q).\tau_s)$} 
 \State $p \leftarrow p + 1$;
 \ElsIf {$(DB(p) < SS(q).\tau_e)$}
 \State $OB \leftarrow true$; $q \leftarrow q + 1$; /* Beat overlapped with {\em bol} */
 \Else\
 \State $OB \leftarrow false$; $q \leftarrow q + 1$; /* No beat overlaps with {\em bol} */
 \EndIf
 \EndWhile
 \State /* Mark the beats */
 \State $i \leftarrow 1$; $j \leftarrow 1$; /* $i$ indexes input $SS$ and $j$ indexes output $MB$ */ 
	\State $MB(1) \leftarrow [SS(1).\beta, SS(1).\tau_s, SS(1).\tau_e, \alpha^{fb}]$; /* Downbeat */
	\State /* Time-stamp of the last $1$-beat as marked */
	\State $last\_beat \leftarrow SS(1).\tau_s$; 
 \While{ $i < length(SS)$}
 \State /* $wide(T) = [T-0.25, T+0.4]$ */
 \If{ $(SS(i+1).\tau_s - last\_beat)$ is within $wide(T)$} 
 \If{ $(SS(i+1).\epsilon = high)$} /* $1$-beat */
 \State $MB(j+1) \leftarrow $
 \State $\quad\quad[SS(i+1).\beta, SS(i+1).\tau_s, SS(i+1).\tau_e, \alpha^{fb}]$;
 \ElsIf {$((SS(i+1).\epsilon = low)$ \& $(OB(i+1) = true))$} 
 \State $MB(j+1) \leftarrow \quad\quad $ /* Stick-beat, $\bot$ */
 \State $\quad\quad[\bot, SS(i+1).\tau_s, SS(i+1).\tau_e, \alpha^{fn}]$; 
 \Else\ /* Error -- undefined beat, $\top$ */
 \State $MB(j+1) \leftarrow [SS(i+1).\beta, SS(i+1).\tau_s, SS(i+1).\tau_e, \top]$;
 \EndIf
 \State $last\_beat \leftarrow SS(i+1).\tau_s$; $i \leftarrow i+1$; $j \leftarrow j+1$;
 \State /* $\frac{1}{2}$-beat; No update to $last\_beat$ */
 \ElsIf {$(SS(i+1).\tau_s - last\_beat) < T$} 
 	\State $MB(j+1) \leftarrow [SS(i+1).\beta, SS(i+1).\tau_s, SS(i+1).\tau_e, \alpha^{hb}]$;
 	\State $i \leftarrow i+1$; $j \leftarrow j+1$;
 \State /* $long\_gap(T) = 2*T-0.25$ */
 \ElsIf {$(SS(i+1).\tau_s  - last\_beat) > long\_gap(T)$} 
 \State /* No beat in a long gap between $SS(i).\tau_s $ \& $SS(i+1).\tau_s$ */
 	\State /* Stick-beat, $\bot$ forced */
 	\State $MB(j+1) \leftarrow [\bot, last\_beat+T, last\_beat+T+0.5, \alpha^{fn}]$; 
 	\State $last\_beat \leftarrow last\_beat+T$; $j \leftarrow j+1$;
 \EndIf
 \EndWhile
 \end{algorithmic}
 \end{footnotesize}
 \end{algorithm}

\subsection{Results and Analysis}
We show examples of beat marking for samples of {\em Joining A} and {\em Joining B Sollukattu}s in Tab.~\ref{tbl:beat_marking}. For {\em Joining A}, the beats are correctly marked in the presence of stick-beat. For {\em Joining B}, $1$- and $\frac{1}{2}$-beats are correctly marked. We also illustrate a case of {\em Sarika Sollukattu} in Tab.~\ref{tbl:beat_marking_sarika}. Here the annotation has 32 beats and the beat marking algorithm could mark only 26 beats. However, the correct match occurred only for 23 beats as in 3 cases a {\em bol} was falsely detected from a stick-beat in the input. This is partly due to segmentation error (hence the beat gets positioned as a $\frac{1}{2}$-beat) and partly due to GMM error. Interestingly, there are 9 cases where the {\em bol} `tei' is correctly recognized, but the beat still could not be marked as the energy of the slices of `tei' are very low. But they have correct positions due to correct recognition of {\em bol}. Hence these get marked as stick-beats and cause lower accuracy.
\begin{table}[!ht]
\centering
\caption{Beat marking on {\em Joining A \& B Sollukattu}s\label{tbl:beat_marking}}
\begin{tabular}{|r|r|l|l||r|r|l|l|} \hline
\multicolumn{4}{|c}{\bf {\em Joining A Sollukattu}} & \multicolumn{4}{||c|}{\bf {\em Joining B Sollukattu}} \\ \hline
\multicolumn{1}{|c}{\bf Start} & \multicolumn{1}{|c}{\bf End} & \multicolumn{1}{|c}{\bf {\em bol}} & \multicolumn{1}{|c}{\bf Beat} & \multicolumn{1}{||c}{\bf Start} & \multicolumn{1}{|c}{\bf End} & \multicolumn{1}{|c}{\bf {\em bol}} & \multicolumn{1}{|c|}{\bf Beat} \\ 
\multicolumn{1}{|c}{\bf Time} & \multicolumn{1}{|c}{\bf Time} & \multicolumn{1}{|c}{\bf {\em }} & \multicolumn{1}{|c}{\bf Info} & \multicolumn{1}{||c}{\bf Time} & \multicolumn{1}{|c}{\bf Time} & \multicolumn{1}{|c}{\bf {\em }} & \multicolumn{1}{|c|}{\bf Info} \\  
\multicolumn{1}{|c}{\bf (sec.)} & \multicolumn{1}{|c}{\bf (sec.)} & \multicolumn{1}{|c}{\bf {\em }} & \multicolumn{1}{|c}{\bf } & \multicolumn{1}{||c}{\bf (sec.)} & \multicolumn{1}{|c}{\bf (sec.)} & \multicolumn{1}{|c}{\bf {\em }} & \multicolumn{1}{|c|}{\bf } \\ \hline \hline
4.36  &  4.72  &  tat  &  B  & 1.04 & 1.39 &  dhit  &  B   \\
5.63  &  6.00  &  dhit  &  B  & 1.97 & 2.32 &  dhit  &  HB   \\
6.85  &  7.41  &  ta  &  B  & 2.90 & 3.26 &  tei  &  B   \\
9.17  &  9.52  &  tat  &  B  & 4.58 & 4.91 &  dhit  &  B   \\
10.34  &  10.70  &  dhit  &  B  & 5.38 & 5.73 &  dhit  &  HB   \\
11.50  &  12.01  &  ta  &  B   & 6.19 & 6.53 &  tei  &  B   \\ \cline{1-4}
\multicolumn{4}{c||}{ } & 7.72 & 8.04 &  dhit  &  B   \\
\multicolumn{4}{c||}{ } & 8.46 & 8.79 &  dhit  &  HB   \\
\multicolumn{4}{c||}{ } & 9.23 & 9.56 &  tei  &  B   \\
\multicolumn{4}{c||}{ } & 10.62 & 10.95 &  dhit  &  B   \\
\multicolumn{4}{c||}{ } & 11.37 & 11.7 &  dhit  &  HB   \\
\multicolumn{4}{c||}{ } & 12.11 & 12.46 &  tei  &  B   \\ \cline{5-8}
\end{tabular}
\end{table}

\begin{table}[!ht]
\centering
\caption{Beat marking on {\em Sarika Sollukattu}\label{tbl:beat_marking_sarika}}
\begin{footnotesize}
\begin{tabular}{|r|r|l|l||r|r|l|l||l|} 
\cline{1-8}
\multicolumn{4}{|c}{\bf Annotation of Beats} & \multicolumn{4}{||c||}{\bf Marking of Beats} \\ \hline
\multicolumn{1}{|c}{\bf Start} & \multicolumn{1}{|c}{\bf End} & \multicolumn{1}{|c}{\bf {\em bol}} & \multicolumn{1}{|c}{\bf Beat} & \multicolumn{1}{||c}{\bf Start} & \multicolumn{1}{|c}{\bf End} & \multicolumn{1}{|c}{\bf {\em bol}} & \multicolumn{1}{|c}{\bf Beat} & \multicolumn{1}{||c|}{\bf Remarks} \\ 
\multicolumn{1}{|c}{\bf Time} & \multicolumn{1}{|c}{\bf Time} & \multicolumn{1}{|c}{\bf {\em }} & \multicolumn{1}{|c}{\bf Info} & \multicolumn{1}{||c}{\bf Time} & \multicolumn{1}{|c}{\bf Time} & \multicolumn{1}{|c}{\bf {\em }} & \multicolumn{1}{|c}{\bf Info} & \multicolumn{1}{||c|}{\bf } \\   
\multicolumn{1}{|c}{\bf (sec.)} & \multicolumn{1}{|c}{\bf (sec.)} & \multicolumn{1}{|c}{\bf {\em }} & \multicolumn{1}{|c}{\bf } & \multicolumn{1}{||c}{\bf (sec.)} & \multicolumn{1}{|c}{\bf (sec.)} & \multicolumn{1}{|c}{\bf {\em }} & \multicolumn{1}{|c}{\bf }  & \multicolumn{1}{||c|}{\bf } \\ \hline \hline
1.94	&	2.45	&	tei	&	B	&	1.81	&	2.17	&	tei	&	B	& Match		\\
2.90	&	3.37	&	a	&	B	&	2.80	&	3.22	&	a	&	B	& Match		\\
3.93	&	4.39	&	tei	&	B	&		    &		    &		&		& No$^a$ `tei' \\
4.91	&	5.33	&	e	&	B	&	4.79	&	5.17	&	e	&	B	& Match		\\
5.86	&	6.35	&	tei	&	B	&	5.75	&	6.15	&	tei	&	B	& Match		\\
6.82	&	7.31	&	a	&	B	&	6.74	&	7.14	&	a	&	B	& Match		\\
7.80	&	8.26	&	tei	&	B	&	 	    &	 	    &	 	&	 	& No$^a$ `tei'	\\
    	&	    	&	  	&	 	&	7.23	&	7.43	&	tat	&	HB	& HB$^b$ `tat' \\
8.74	&	9.17	&	e	&	B	&	8.63	&	9.00	&	e	&	B	& Match		\\
9.68	&	10.20	&	tei	&	B	&	9.59	&	9.96	&	tei	&	B	& Match		\\
10.64	&	11.07	&	a	&	B	&	10.54	&	10.93	&	a	&	B	& Match		\\
11.48	&	11.91	&	tei	&	B	&	 	    &	 	    &	 	&	 	& No$^a$ `tei'	\\
    	&	    	&	 	&	 	&	10.97	&	11.26	&	tat	&	HB	& HB$^b$ `tat' \\
12.37	&	12.79	&	e	&	B	&	12.25	&	12.61	&	e	&	B	& Match		\\
13.30	&	13.75	&	tei	&	B	&		    &		    &		&		& No$^a$ `tei' \\
14.19	&	14.62	&	a	&	B	&	14.07	&	14.46	&	a	&	B	& Match		\\
15.14	&	15.59	&	tei	&	B	&		    &		    &		&		& No$^a$ `tei' \\
16.00	&	16.49	&	e	&	B	&	15.90	&	16.25	&	e	&	B	& Match		\\
16.95	&	17.37	&	tei	&	B	&	16.84	&	17.24	&	tei	&	B	& Match		\\
17.93	&	18.33	&	a	&	B	&	17.78	&	18.16	&	a	&	B	& Match		\\
18.86	&	19.24	&	tei	&	B	&		    &		    &		&		& No$^a$ `tei'	\\
19.77	&	20.18	&	e	&	B	&	19.65	&	20.01	&	e	&	B	& Match		\\
20.64	&	21.11	&	tei	&	B	&	20.54	&	20.92	&	tei	&	B	& Match		\\
21.52	&	21.96	&	a	&	B	&	21.44	&	21.85	&	a	&	B	& Match		\\
22.43	&	22.84	&	tei	&	B	&		    &		    &		&		& No$^a$ `tei' \\
23.26	&	23.68	&	e	&	B	&	23.17	&	23.51	&	e	&	B	& Match		\\
24.17	&	24.61	&	tei	&	B	&	24.05	&	24.44	&	tei	&	B	& Match		\\
25.02	&	25.46	&	a	&	B	&	24.92	&	25.32	&	a	&	B	& Match		\\
25.90	&	26.32	&	tei	&	B	&	    	&	    	&	 	&	 	& No$^a$ `tei'	\\
    	&	    	&	 	&	 	&	25.34	&	25.60	&	na	&	HB	& HB$^c$ `na' \\
26.74	&	27.16	&	e	&	B	&	26.63	&	27.00	&	e	&	B	& Match		\\
27.65	&	28.17	&	tei	&	B	&	27.53	&	27.93	&	tei	&	B	& Match		\\
28.60	&	29.22	&	a	&	B	&	28.48	&	28.86	&	a	&	B	& Match		\\
29.52	&	29.97	&	tei	&	B	&		    &		    &		&		& No$^a$ `tei' 	\\
30.41	&	30.83	&	e	&	B	&	30.30	&	30.64	&	e	&	B	& Match		\\
 \hline
 \multicolumn{9}{l}{ } \\
\multicolumn{9}{p{7cm}}{$\bullet$ Every correct match of time, {\em bol} \& event is marked  `Match'} \\
\multicolumn{9}{p{7cm}}{$\bullet$ $^a$: `tei' is correctly detected but marked as stick-beat due to very low energy of the `tei' slice and hence skipped} \\
\multicolumn{9}{p{7cm}}{$\bullet$ $^b$: `tat' is wrongly detected (from $\bot$) and marked as HB} \\
\multicolumn{9}{p{7cm}}{$\bullet$ $^c$: `na' is wrongly detected (from $\bot$) and marked as HB} \\
\multicolumn{9}{p{7cm}}{$\bullet$ \# of beats in annotation = 32. \# correctly matched = 23. Accuracy = 71.88\%} \\
\end{tabular}
\end{footnotesize}
\end{table}

To check for the overall accuracy of beat marking, we compare it against the annotations of beat information, time-stamp, and {\em bol} for SR1 data set (Tab.~\ref{tbl:Sollukattu_Data_Set}). Consider an audio signal $f^s(t)$. Let $b \in MB$ be a marked beat where $b = [\tau_s, \tau_e, \beta(bol), \alpha]$, $\alpha \in \{\alpha^{fb}, \alpha^{hb}, \alpha^{fn}\}$, and $MB$ is computed from $f^s(t)$ by Algorithm~\ref{algo_bm}. Also, let $a \in AB$ be a beat in the annotation $AB$ of $f^s(t)$ and is represented in the same way. Now we compute the match between $MB$ and $AB$ based on three parameters:

\begin{enumerate}
\item {\bf Time Match}: $b \in MB$ matches $a \in AB$ if they overlap in time. That is, $[b.\tau_s, b.\tau_e] \cap [a.\tau_s, a.\tau_e] \neq \phi$. If $u$ out of $|AB|$ beats match, we have $u/|AB|*100$\% time match.

\item {\bf {\em bol} Match}: If $b \in MB$ matches $a \in AB$ in time, we check if their {\em bol}s agree. That is, $b.\beta(bol) = a.\beta(bol)$. If $v$ out of $|AB|$ {\em bol}s match, we have $v/|AB|*100$\% {\em bol} match.

\item {\bf Event / Beat Info Match}: If $b \in MB$ matches $a \in AB$ in time, we check if their events match. That is, $b.\alpha = a.\alpha$. If $w$ out of $|AB|$ events match, we have $w/|AB|*100$\% event match.
\end{enumerate}

In Tab.~\ref{tbl:Beat_Marking_Results}, we have computed the matches in two sets -- first using only $1$-beats and then using $1$- as well as $\frac{1}{2}$-beats. These have been done for SR1 data set. {\bf Using $1$-beats we achieve 94.46\%, 91.83\%, and	90.72\%	accuracy for time, {\em bol}, and event matches respectively. Using $1$- as well as $\frac{1}{2}$-beats, however, the accuracy drops by 5\%--10\% to 88.17\%, 81.88\%, and 84.75\% respectively}. This drop is due to less robust estimation of the time-stamps of $\frac{1}{2}$-beats.

\begin{table}[!ht]
\centering
\caption{Results of beat marking\label{tbl:Beat_Marking_Results}}
\renewcommand{\arraystretch}{1.2}
\centering
\begin{scriptsize}
\begin{tabular}{|l|r|r|r||r|r|r|}
\cline{2-7}
\multicolumn{1}{c|}{} & \multicolumn{6}{c|}{\bf Percentage of Match in Beat Marking} \\ \cline{2-7}
\multicolumn{1}{c|}{ } & \multicolumn{3}{c||}{\textbf{For $1$-beats}}  & \multicolumn{3}{c|}{\textbf{For $1$- \& $\frac{1}{2}$-beats}}  \\ \hline
\multicolumn{1}{|l|}{\textbf{\em Sollakattu}} & \multicolumn{1}{c|}{\textbf{Time}} & \multicolumn{1}{c|}{\textbf{\em Bol}} & \multicolumn{1}{c||}{\textbf{Event}} & \multicolumn{1}{c|}{\textbf{Time}} & \multicolumn{1}{c|}{\textbf{\em Bol}} & \multicolumn{1}{c|}{\textbf{Event}} \\ \hline \hline
Joining A	&	100.00	&	100.00	&	100.00	&	100.00	&	100.00	&	100.00	\\ \hline
Joining B	&	100.00	&	100.00	&	100.00	&	100.00	&	100.00	&	100.00	\\ \hline
Joining C	&	100.00	&	100.00	&	100.00	&	100.00	&	100.00	&	100.00	\\ \hline
Kartari Utsanga	&	66.67	&	66.67	&	66.67	&	60.42	&	37.50	&	45.83	\\ \cline{2-7}
Mandi Sarikkal	&	89.58	&	77.08	&	35.42	&	60.42	&	44.79	&	37.50	\\ \hline
Kuditta Mettu 1	&	100.00	&	100.00	&	100.00	&	100.00	&	100.00	&	100.00	\\ \cline{2-7}
Kuditta Mettu 2	&	100.00	&	100.00	&	100.00	&	100.00	&	100.00	&	100.00	\\ \hline
Kuditta Nattal A	&	54.17	&	54.17	&	54.17	&	54.17	&	54.17	&	54.17	\\ \hline
Kuditta Nattal B	&	81.25	&	81.25	&	81.25	&	87.50	&	87.50	&	87.50	\\ \hline
Kuditta Tattal 1	&	100.00	&	100.00	&	100.00	&	100.00	&	100.00	&	100.00	\\ \cline{2-7}
Kuditta Tattal 2	&	100.00	&	97.50	&	100.00	&	100.00	&	97.50	&	100.00	\\ \cline{2-7}
Kuditta Tattal 3	&	100.00	&	100.00	&	100.00	&	100.00	&	100.00	&	100.00	\\ \cline{2-7}
Kuditta Tattal 4	&	100.00	&	87.50	&	100.00	&	100.00	&	87.50	&	100.00	\\ \hline
Natta 1	&	100.00	&	100.00	&	100.00	&	96.43	&	96.43	&	96.43	\\ \cline{2-7}
Natta 2	&	100.00	&	100.00	&	100.00	&	100.00	&	100.00	&	100.00	\\ \hline
Paikkal	&	100.00	&	100.00	&	100.00	&	75.00	&	75.00	&	75.00	\\ \hline
Pakka 1	&	100.00	&	100.00	&	100.00	&	100.00	&	100.00	&	100.00	\\ \cline{2-7}
Pakka 2	&	98.44	&	98.44	&	98.44	&	98.44	&	98.44	&	98.44	\\ \hline
Sarika	&	71.88	&	71.88	&	71.88	&	71.88	&	71.88	&	71.88	\\ \hline
Tatta A	&	100.00	&	100.00	&	100.00	&	100.00	&	75.00	&	100.00	\\ \hline
Tatta B	&	100.00	&	100.00	&	100.00	&	100.00	&	100.00	&	100.00	\\ \hline
Tatta C	&	100.00	&	100.00	&	100.00	&	100.00	&	100.00	&	100.00	\\ \hline
Tatta D	&	100.00	&	100.00	&	100.00	&	100.00	&	100.00	&	100.00	\\ \hline
Tatta E	&	100.00	&	91.67	&	100.00	&	100.00	&	91.67	&	100.00	\\ \hline
Tatta F	&	100.00	&	92.86	&	100.00	&	100.00	&	92.86	&	100.00	\\ \hline
Tatta G	&	100.00	&	100.00	&	100.00	&	100.00	&	100.00	&	100.00	\\ \hline
Tei Tei Dhatta	&	100.00	&	100.00	&	100.00	&	56.25	&	53.13	&	53.13	\\ \hline
Tirmana A	&	100.00	&	60.00	&	100.00	&	78.57	&	42.86	&	78.57	\\ \hline
Tirmana B	&	95.83	&	95.83	&	95.83	&	97.62	&	92.86	&	97.62	\\ \hline
Tirmana C	&	91.67	&	79.17	&	87.50	&	92.50	&	52.50	&	87.50	\\ \hline \hline
													
{\bf Cumulative}	&	{\bf 94.46}	&	{\bf 91.83}	&	{\bf 90.72}	&	{\bf 88.17}	&	{\bf 81.88}	&	{\bf 84.75}	\\ \hline
\multicolumn{7}{c}{ } \\
\multicolumn{7}{p{6cm}}{$\bullet$ Results for SR1 data set (Tab.~\ref{tbl:Sollukattu_Data_Set})} \\
\multicolumn{7}{p{7cm}}{$\bullet$ Multiple samples from the same {\em Sollukattu} are serially numbered} \\
\end{tabular}
\end{scriptsize}
\end{table}

\section{Conclusions\label{sec:conclusion}}
In this paper, we first detect the $1$-beats from the onset envelope of the signal by using algorithms from~\cite{mallick2018characterization}. We then apply speech processing techniques for {\em Sollukattu} recognition as it is a mixture of vocal and instrumental music. We also estimate the tempo period from the signal and generate a complete annotation of the audio signal by beat marking. {\bf We achieve 85\% accuracy in {\em bol} recognition, 95\% in {\em Sollukattu} recognition, 96\% in tempo period estimation, and over 90\% in beat marking. The proposed scheme offers a simple but effective approach to fully structurally analyze the music of an Indian Classical Dance form.}

The algorithms developed in this paper can be used in many applications including:
\begin{itemize}
\item {\bf Automatic Audio Annotation}: Our algorithm generates automatic annotation of {\em Bharatanatyam Adavu} from the accompanying audio. The audio events are detected and specified at multiple levels of granularity (Tab.~\ref{tbl:beat_marking}).
\item {\bf Dance Video Segmentation}: Dance video segmentation is a challenging task. The researchers often do not attempt the problem and develop their video solutions on the pre-segmented data. The algorithms developed here can help to segments the video based on the inherent structure of the {\em Adavu}s, as they are driven by the music. We use these annotations for video segmentation in~\cite{mallick2018characterization} and {\em Adavu} recognition in~\cite{mallick2017Adavu}.   
\item {\bf New {\em Sollukattu} Annotation}: The algorithms work on dictionary based speech recognition. Hence, new {\em Sollukattu}s can be recognized by just adding {\em bol}s in the dictionary and training appropriately. 
\end{itemize}


\section*{Acknowledgment}
The work of the first author is supported by TCS Research Scholar Program of Tata Consultancy Services of India.

{\small
\bibliographystyle{plain}
\bibliography{References,BibAudio}
}

\end{document}